\documentclass[longauth]{aa}  

\usepackage{graphicx}
\usepackage{txfonts}
\usepackage{hyperref}
\usepackage{natbib}
\usepackage{lscape}
\usepackage{pdflscape}
\usepackage{soul}

\hypersetup{
        colorlinks=true,    
        linkcolor=blue,  
        citecolor=blue,    
        filecolor=cyan,     
        urlcolor=black,      
} 

\usepackage{color}

\begin{document} 

\title{A photo-$z$ cautionary tale: Redshift confirmation of COSBO-7 at $z=2.625$}

\titlerunning{A cautionary tale of photo-$z$}
\authorrunning{Jin et al.}

\author{
Shuowen Jin\inst{1,2,\thanks{Marie Curie Fellow}},
Nikolaj B. Sillassen\inst{1,2},
Jacqueline Hodge\inst{3},
Georgios E. Magdis\inst{1,2,4},
Francesca Rizzo\inst{5,1}, 
Caitlin Casey\inst{6,1},
Anton M. Koekemoer\inst{7},
Francesco Valentino\inst{8},
Vasily Kokorev\inst{6},
Benjamin Magnelli\inst{9},
Raphael Gobat\inst{10},
Steven Gillman\inst{1,2},
Maximilien Franco\inst{6},
Andreas Faisst\inst{11},
Jeyhan Kartaltepe\inst{12},
Eva Schinnerer\inst{13},
Sune Toft\inst{1,4}, 
Hiddo S. B. Algera\inst{14,15},
Santosh Harish\inst{10}, 
Minju Lee\inst{1,2}, 
Daizhong Liu\inst{16},  
Marko Shuntov\inst{1,4}, 
Margherita Talia\inst{17,18}, 
Aswin Vijayan\inst{1,2}
          }

   \institute{Cosmic Dawn Center (DAWN), Denmark\\
      \email{shuji@dtu.dk}
    \and
            DTU Space, Technical University of Denmark, Elektrovej 327, DK-2800 Kgs. Lyngby, Denmark
    \and
            Leiden Observatory, Leiden University, P.O. Box 9513, 2300 RA Leiden, The Netherlands
    \and
            Niels Bohr Institute, University of Copenhagen, Jagtvej 128, DK-2200 Copenhagen, Denmark
    \and
            Kapteyn Astronomical Institute, University of Groningen, 9700 AV Groningen, The Netherlands
    \and
            The University of Texas at Austin, 2515 Speedway Blvd Stop C1400, Austin, TX 78712, USA
    \and
            Space Telescope Science Institute, 3700 San Martin Drive, Baltimore, MD 21218, USA
    \and
            European Southern Observatory (ESO), Karl-Schwarzschild-Strasse 2, Garching 85748, Germany
    \and
            Université Paris-Saclay, Université Paris Cité, CEA, CNRS, AIM, 91191 Gif-sur-Yvette, France
    \and
            Instituto de Física, Pontificia Universidad Católica de Valparaíso, Casilla 4059, Valparaíso, Chile
    \and
            Caltech/IPAC, MS 314-6, 1200 E. California Blvd. Pasadena, CA 91125, USA
    \and
            Laboratory for Multiwavelength Astrophysics, School of Physics and Astronomy, Rochester Institute of Technology, 84 Lomb Memorial Drive, Rochester, NY 14623, USA
    \and
            Max-Planck-Institut für Astronomie, Königstuhl 17, 69117 Heidelberg, Germany
    \and
            Hiroshima Astrophysical Science Center, Hiroshima University, 1-3-1 Kagamiyama, Higashi-Hiroshima, Hiroshima 739-8526, Japan
    \and
            National Astronomical Observatory of Japan, 2-21-1, Osawa, Mitaka, Tokyo, Japan
    \and    
            Purple Mountain Observatory, Chinese Academy of Sciences, 10 Yuanhua Road, Nanjing 210023, China
    \and
            University of Bologna, Department of Physics and Astronomy (DIFA), Via Gobetti 93/2, I-40129, Bologna, Italy
    \and
            INAF - Osservatorio di Astrofisica e Scienza dello Spazio, via Gobetti 93/3 - 40129, Bologna - Italy
            }
   \date{Received XXX / Accepted XXX}

\abstract
{Photometric redshifts are widely used in studies of dusty star-forming galaxies (DSFGs), but catastrophic photo-$z$ failure can undermine all redshift-dependent results. Here we report the spectroscopic redshift confirmation of COSBO-7, a strongly lensed DSFG in the COSMOS-PRIMER field. Recently, a photometric redshift solution of $z\gtrsim7.0$ was reported for COSBO-7 based on ten bands of {\it James Webb} Space Telescope (JWST) NIRCam and MIRI imaging data. This $z$ value was favored by four independent spectral energy distribution (SED) fitting codes, and the result provided an appealing candidate for the most distant massive DSFG known to date. This photo-$z$ solution was also supported by a single line detection in Atacama Large Millimeter Array (ALMA) Band 3 consistent with CO(7-6) at $z=7.46$. However, our new ALMA observations robustly detect two lines in Band 6 identified as CO(7-6) and [CI](2-1) at $z_{\rm spec}=2.625$, and thus the Band 3 line as CO(3-2). These three robust line detections decidedly place COSBO-7  at $z=2.625$, refuting the photo-$z$ solution. We derive physical parameters by fitting near-infrared(NIR)-to-millimeter(mm) photometry and lens modeling, revealing that COSBO-7 is a main sequence galaxy. We examine possible reasons for this photo-$z$ failure and attribute it to (1) the likely underestimation of photometric uncertainties at 0.9\,$\mu$m and 1.15 \,$\mu$m; and (2) the lack of photometry at wavelengths beyond 20\,$\mu$m. Notably, we recover a bona fide $z_{\rm phot}\sim 2.3$ by including the existing MIPS $24\,\mu$m photometry, demonstrating the critical importance of mid-infrared (MIR) data in bolstering photo-$z$ measurements. This work highlights a common challenge in modeling the SEDs of DSFGs, and provides a cautionary tale regarding the reliability of photometric redshifts as well as pseudo-spectroscopic redshifts based on single line detection.
}


\maketitle


\section{Introduction}

Redshift is one of the most important parameters of a galaxy, and determining the redshift of a galaxy is the first and most critical step in revealing its nature.
To date, the most distant galaxies have been confirmed at $z\sim14$ \citep{Carniani2024z14} by spectroscopy from the {\it James Webb} Space Telescope (JWST).
For dusty star-forming galaxies (DSFGs), given the severe dust attenuation and extreme faintness in optical and near-infrared (NIR) wavelengths (e.g., \citealt{Wang2019Natur,Smail2021Kdrop}), detecting molecular and neutral line features in millimeter (mm) and submillimeter(submm) wavelengths is a more efficient way of confirming their redshifts. Thanks to the advanced mm and submm interferometers, such as Atacama Large Millimeter Array (ALMA) and Northern Extended Millimeter Array (NOEMA), distant DSFGs have been spectroscopically confirmed at $z>5$ and out to the epoch of reionization \citep[EoR, $z\sim7$; e.g.,][]{Walter2012, Vieira2010,Vieira2013,Riechers2013Nature, Riechers2017,Strandet2017,Zavala2017, Marrone2017Nature,Endsley2022z68,Fudamoto2021Nature,Hygate2023z7, Rowland2024z7}, proving these facilities to be a powerful ``redshift machine'' \citep{Vieira2013,Neri2020noema,ChenCC2022alma,Cox2023zGal}.

However, performing spectroscopy on DSFGs remains observationally expensive, which limits the size and completeness of current spec-$z$ samples. Alternatively, photometric redshifts, which are estimated by modeling spectral energy distributions (SEDs)
with optical and NIR photometry (e.g., \citealt{Arnouts1999LePhare,Ilbert2006LePhare,da_Cunha2008magphys,Brammer2008EAZY,Kriek2009FAST,Carnall2018Bagpipes,Boquien2019CIGALE}), are widely used and dominate in literature studies of DSFGs (e.g., \citealt{Boone2011,Wardlow2011,Simpson2014,Miettinen2015,Wang2019Natur,Gomez-Guijarro2019,Dudzeviciute2020SMG,Smail2021Kdrop}).
In comparison to dust-free galaxies, constraining the redshifts of DSFGs is particularly challenging because of severe dust attenuation. 
Nowadays, the situation has been dramatically improved with the data from JWST. With its unprecedented sensitivity and long-wavelength coverage, photometric redshifts have been estimated out to $z>10-16$ for dust-free galaxies (e.g., \citealt{Naidu2022photoz,Finkelstein2022z12,Harikane2023z10,Atek2023z16,Casey2024z10,Chakraborty2024z11}) and $z\sim8$ for DSFGs (e.g., \citealt{Barrufet2023dark,Akins2023z8candi}).
Nevertheless, photometric redshifts can still fail catastrophically ---that is, they can be  catastrophically erroneous--- for dust-free galaxies, even with multiband JWST photometry. For example, the galaxy CEERS-93316 was reported with a photometric redshift of $z\sim16.4$ based on a SED fitting with seven bands of JWST NIRCam photometry \citep{Donnan2023z16}, but was eventually confirmed to be a $z=4.9$ dusty galaxy based on JWST NIRSpec spectroscopy \citep{Arrabal_Haro2023Natur}, demonstrating that dusty starbursts can masquerade as ultrahigh-redshift galaxies \citep{Zavala2023photoz,Naidu2022photoz}. As the misidentification of photo-$z$ can undermine all results that are dependent on redshift, it is vital to examine whether such photo-$z$ failure can also happen for DSFGs preselected from mm and submm surveys.

\begin{figure*}[ht]
\centering
\includegraphics[width=0.98\textwidth]{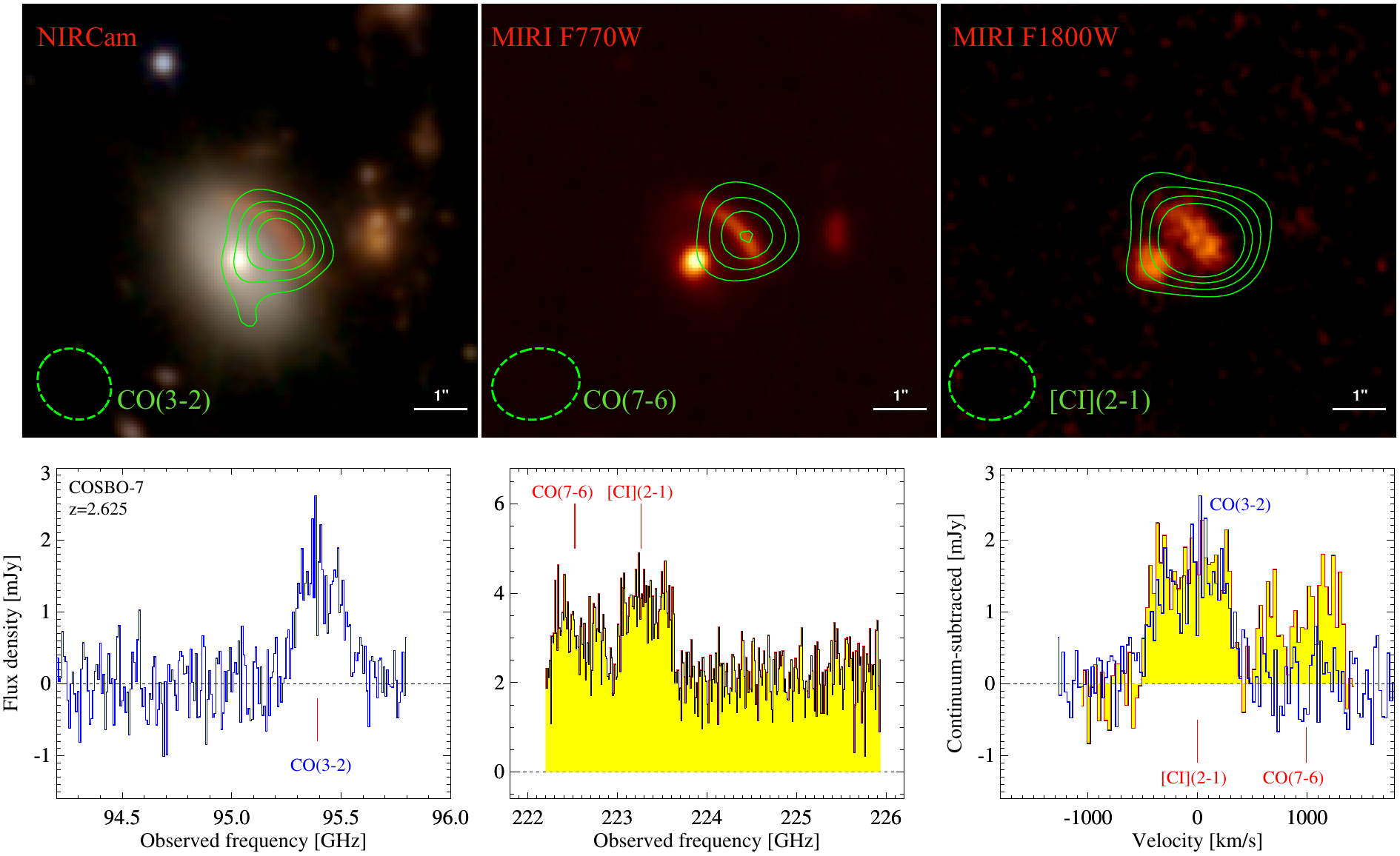}
\caption{JWST images and ALMA spectra of COSBO-7. {\it Top:}  NIRCam color image (Blue: F090W+F115W+F150W; Green: F200W+F277W; Red: F356W+F410M+F444W), and MIRI images overlaid with contours of CO and [CI] emission. Contours are shown at 4, 6, 8, and 10$\sigma$ levels. The beams are shown as dashed ellipses. {\it Bottom:}  Left and middle panels show the CO(3-2) and CO(7-6)+[CI](2-1) spectra in observed frequencies. The right panel shows the continuum-subtracted spectra as a function of velocity.}
\label{fig-spec}
\end{figure*}

Recently, a $z>7$ DSFG candidate, COSBO-7, was reported by \cite{Ling2024cosbo7} based on exquisite imaging data from JWST. 
\cite{Ling2024cosbo7} performed extensive imaging fitting and SED analysis using the JWST NIRCam and MIRI data, and found that COSBO-7 is not detected in NIRCam/F090W but is well detected in nine bands from NIRCam/F115W to MIRI/F1800W. Using the photometry measured on the lens-subtracted images, these authors calculated photometric redshifts of $z_{\rm phot}=6.9-7.7$, finding that four SED algorithms, namely \texttt{LePhare}
\citep{Ilbert2006LePhare}, \texttt{EAZY} \citep{Brammer2008EAZY}, \texttt{Bagpipes}
\citep{Carnall2018Bagpipes}, and \texttt{CIGALE} \citep{Boquien2019CIGALE} agree on a best-fit solution of $z\sim7.0$. This makes COSBO-7 an appealing candidate for the most distant DSFG to date. 
Furthermore, \cite{Ling2024cosbo7} reported a line detection at 95.4~GHz in ALMA archival data that would be consistent with CO(7-6) emission at $z=7.46$, that is, at a redshift very close to the $z_{\rm phot}$ solution.
Nevertheless, robust spectroscopic confirmation with multiple lines was still missing.

In this Letter, we report unambiguous spectroscopic redshift confirmation of COSBO-7 and discuss the implications of erroneous photometric redshift estimates and the lessons learned from this cautionary tale.
We adopt a flat $\Lambda$CDM cosmology with $H_0=70$~km~s$^{-1}$~Mpc$^{-1}$ and $\Omega_M=0.27$, and a Chabrier initial mass function \citep{Chabrier2003}.

\section{Selection and data}

\subsection{Selection}

COSBO-7 (Right Ascension (RA)=10:00:23.97, Declination (Dec.)=+02:17:50.0) was originally discovered in a flux-limited IRAM/MAMBO-2 1.2mm imaging survey by \cite{Bertoldi2007}. It is one of the brightest submm sources in the COSMOS field, and is also detected in the AzTEC and SCUBA-2 surveys ($S_{\rm 1mm}\sim2$~mJy, $S_{\rm 850\mu m}\sim10$~mJy; \citealt{Aretxaga2011,Geach2017scuba2,Simpson2019SCUBA2}).
COSBO-7 is not detected in deep HST images of the COSMOS-CANDELS field, indicating an extremely dust-obscured nature.
A secure counterpart of COSBO-7 was first identified in radio wavelengths at VLA 1.4~GHz and 3~GHz \citep{Schinnerer2010,Smolcic2017}.
COSBO-7 was identified as a lensing system by \cite{Jin2018cosmos}, wherein a high-z submillimeter galaxy is lensed by a foreground elliptical galaxy at $z_{\rm spec}=0.36$. 
COSBO-7 was observed with ALMA Band 7 in the project 2016.1.00463.S
(PI: Y. Matsuda) and was cataloged by \cite{Simpson2020alma}.
Recently, COSBO-7 was observed with JWST/NIRCam and MIRI as part of the PRIMER survey \citep{Dunlop2021PRIMER}.
The MIRI image clearly reveals a lensing arc in the MIRI 7.7$\mu$m band while a counter-image is found on the ALMA 870$\,\mu$m map \citep{Pearson2024cosbo7}, confirming the strong lensing nature of the system. 
\cite{Ling2024cosbo7} performed an extensive photometric analysis of COSBO-7 
by exploiting the JWST imaging data from NIRCam F090W to MIRI F1800W band after subtracting the foreground lens. These authors found that COSBO-7 remains undetected in F090W, but is well detected in nine bands from F115W to F1800W. With the JWST photometry, \cite{Ling2024cosbo7} performed SED fitting using four SED codes that all converged to a photo-$z$ solution of $z\gtrsim7.0$. 

\subsection{ALMA}

The first spectroscopic follow-up of COSBO-7 was carried out with ALMA Band 3 line scans in Cycle 9 (ID:2022.1.00863.S; PI: J. Hodge) as part of a redshift scan program investigating ten radio-selected, optically dark DSFGs. A strong line was detected at 95.4~GHz (Fig.~\ref{fig-spec}, bottom-left); however the single line detection was insufficient to confirm the redshift of the source. Driven by the photometric redshift $z\gtrsim7.0$ by \cite{Ling2024cosbo7}, it was reasonable to postulate that the 95.4~GHz line originates from  CO(7-6) emission at $z=7.458$. Further, the $z=7.458$ solution was also supported by a multitude of indirect evidence: (1) a tentative line at 95.7~GHz, consistent with [CI](2-1) emission at $z=7.458$ ; (2) F410M excess, indicative of  [OIII]+H$\beta$ emission at $z\sim7.4$; (3) MIPS 24$\mu m$ excess that is consistent with a 3.3$\mu$m polycyclic aromatic hydrocarbon (PAH) feature at $z=7.458$; and (4) a well-fitted panchromatic SED from NIR to radio wavelengths. However, without a robust detection of a second line, the redshift of the source remained ambiguous. Consequently, we proposed 0.5 hr of  ALMA Band 6 observations through DDT time, aiming to detect [CII]158$\mu$m and decidedly determine the redshift of COSBO-7.

The DDT program (ID: 2023.A.00021.S; PI: S. Jin) was approved and the observation was executed on March 21 2024 in C-1 configuration. The frequency tuning covers 222.18--225.94 GHz for lines and 235.60--239.48 GHz for continuum. The on-source integration is 15 mins, and self-calibration was performed. This gives a rms sensitivity of 0.107 mJy/beam per 500~km~s$^{-1}$, and a beam size of $1.43''\times1.14''$ with natural weighting.

The raw data of the ALMA programs mentioned above were reduced and calibrated using the standard ALMA CASA pipeline \citep{McMullin2007CASA}. Following our established pipeline from \cite{Jin2019alma,Jin2022}, we converted the calibrated measurement sets to \texttt{uvfits} format for further analysis in $uv$ space with the GILDAS software. The 1D spectrum was extracted using the GILDAS \texttt{uvfit} routine on the $uv$ tables at all frequencies, where we adopted a point-source model on the fixed position of the ALMA continuum peak.
The continuum and line maps are cleaned using the GILDAS HOGBOM clean routine.
Given that COSBO-7 is resolved in the ALMA data, we measured the continuum and integrated line fluxes on the clean images using an aperture of $r\sim2.5''$, which maximizes the integrated S/Ns.
We also measured the photometry in ALMA Band 4 (ID: 2021.1.00705.S; PI: O. Cooper), and adopted 870$\mu$m photometry in the A3COSMOS catalog \citep{Liu2019A3COSMOS} measured from Band7 data (ID: 2016.1.00463.S,
PI: Y. Matsuda). We list the line fluxes in Table~\ref{tab:physpars} and continuum fluxes in Table~\ref{tab:photo}.

\subsection{JWST}

COSBO-7 was observed with JWST NIRCam and MIRI in ten bands: F090W, F115W, F150W, F200W, F277W, F356W, F410M, and F444W in NIR, and F770W and F1800W in mid-infrared (MIR). \cite{Ling2024cosbo7} modeled the foreground lens using \texttt{Galfit} and measured the photometry on the lens-subtracted residual images. The lens is well modeled, and so the counter image of the arc is recovered in the residual map. The photometry was carefully measured on the PSF-matched and aperture-matched residual images, and therefore we directly adopt the photometry from \cite{Ling2024cosbo7}. As \cite{Ling2024cosbo7} measured $2\sigma$ upper limits of F090W using two different apertures, we adopted the flux limit measured within the larger aperture, but use the $3\sigma$ limit in this work for reasons discussed in Sect.~\ref{discuss1}. For visualization and lensing modeling, we used the PRIMER mosaics produced by M. Franco and S. Harish from the COSMOS-Web team \citep{Casey2023cosmosweb}.

\subsection{Ancillary data}

The FIR and radio photometry of COSBO-7 were already measured in the COSMOS Super-deblended catalog \citep{Jin2018cosmos}. However, the Herschel photometry is noisy because too many priors were fitted within the Herschel beams; that is, it exhibits a high level of crowding. The ALMA data show that COSBO-7 is the only submm-emitting source within the ALMA Band 7 ($r<8''$) and Band 3 primary beam ($r<30''$), indicating negligible blending and no contributions from neighboring sources. Hence, we reran our ``super-deblending'' pipeline with improved priors on Herschel maps as done by \cite{Sillassen2024NICE}, assuming that COSBO-7 is the only source contributing to the Herschel fluxes. As listed in Table~\ref{tab:photo}, the newly measured Herschel photometry shows solid detection in the PACS 160$\mu$m and the SPIRE bands (Fig.~\ref{fig:sed}).

\section{Results}
\subsection{Redshift confirmation}

As shown in the bottom-middle panel of Fig.~\ref{fig-spec}, the DDT program did not detect any line at the expected frequency of 224.7~GHz of [CII] at $z=7.458$. 
Instead, and quite surprisingly, two lines are solidly detected at 222.52~GHz and 223.27~GHz with S/N$=11$ and 23, respectively (Fig.~\ref{fig-spec}). The two lines perfectly match the CO(7-6) and [CI](2-1) transitions at $z=2.625$. Moreover, the 95.4~GHz line is also fitted at the exact frequency of $z=2.625$ CO(3-2). Further, the line widths of the [CI] and CO lines are also consistent with a full width at zero intensity FWZI$=$850~km~s$^{-1}$ (Fig.~\ref{fig-spec}, bottom-right). Therefore, the three solid line detections unambiguously confirm the redshift of COSBO-7 to $z_{\rm spec}=2.625$ instead of $z_{\rm phot}\gtrsim7.0$.

\begin{figure}%
\centering
\includegraphics[width=0.48\textwidth]{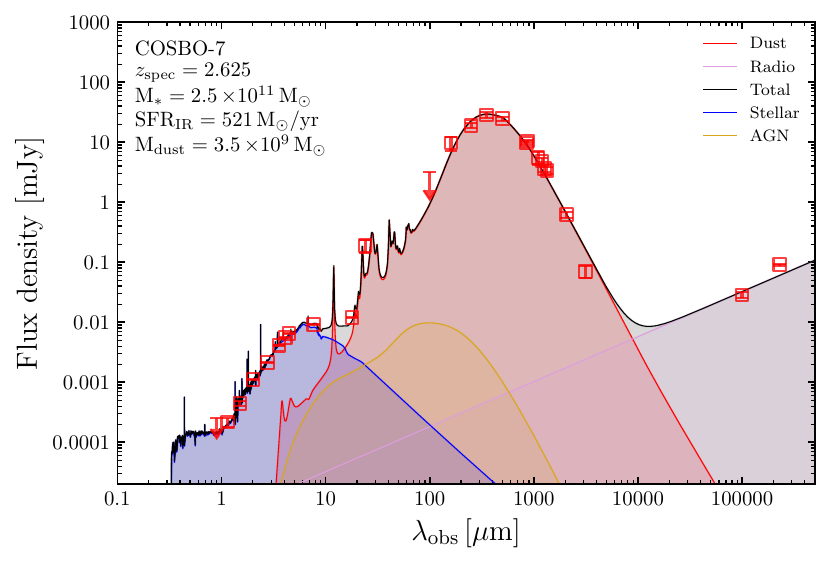}
\caption{Panchromatic SED of COSBO-7 fitted with \texttt{STARDUST} \citep{Kokorev2021stardust}. The F090W upper limit is shown at the $3\sigma$ level. Radio photometry is not included in the fitting; we extrapolated a radio component using the IR luminosity and the IR--radio relation from \cite{Delvecchio2021}. Parameters are not corrected for lensing magnification.
}
\label{fig:sed}
\end{figure}

\begin{table}
    \caption{Physical properties of COSBO-7}
    \centering
    \setlength{\tabcolsep}{2pt}
    \renewcommand\arraystretch{1.5}
    \begin{tabular}{c c c c c c c c}
    \hline\hline
        ID  & COSBO-7 \\
        RA  & 10:00:23.97\\
        Dec & +02:17:50.0 \\
        $z$ & $2.6250\pm0.0007$\\
        $\mu$ & $3.6^{+2.0}_{-0.9}$\\
        \hline
        $I_{\rm CO(3-2)}$ [Jy~km~s$^{-1}$] &   $1.93\pm0.11$ \\
        $I_{\rm CO(7-6)}$ [Jy~km~s$^{-1}$] &   $0.88\pm0.08$ \\
        $I_{\rm [CI](2-1)}$ [Jy~km~s$^{-1}$] &   $1.80\pm0.08$ \\
        \hline
        $A_V$ [mag] &  $1.95\pm0.01$ \\
        $M_\ast$ [$10^{11}{\rm M_\odot}$] & $2.50\pm0.23$ \\
        ${\rm SFR}_{\rm IR}$ [${\rm M_\odot\,yr^{-1}}$] & $ 521_{-45}^{+122}$\\
        $M_{\rm gas,[CI]}$ [${\rm 10^{11} M_\odot}$] & $5.04\pm1.70$\\
        ${\rm SFE}$ [${\rm Gyr^{-1}}$] & $1.03\pm0.35$\\
        \hline
        $T_{\rm dust,thick}$ [K] & $36.5^{+0.6}_{-0.6}$\\
        $M_{\rm dust,thick}$ [$10^{9}{\rm M_\odot}$] & $2.9^{+0.3}_{-0.3}$ \\
        \hline
        $T_{\rm dust,thin}$ [K] & $25.9^{+1.0}_{-0.9}$\\
        $M_{\rm dust,thin}$ [$10^{9}{\rm M_\odot}$] & $5.3^{+0.9}_{-0.7}$ \\
        \hline\hline
    \end{tabular}
    {\\Notes: These parameters are not corrected for magnification $\mu$.}
    \label{tab:physpars}
\end{table}

\subsection{Physical properties and lensing model}

With the confirmed $z_{\rm spec}=2.625$, we derive some physical parameters of COSBO-7 by SED fitting with multiwavelength photometry. As shown in Fig.~\ref{fig:sed}, the SED is well fitted from NIR to radio wavelengths using \texttt{Stardust} \citep{Kokorev2021stardust}. Specifically, the MIPS 24$\mu$m excess is well fitted by strong PAH features at a rest-frame of $6-8\mu$m, while the MIR AGN contribution to the total IR luminosity is negligible ($<2\%$). We report the best-fit parameters in Fig.~\ref{fig:sed} and Table~\ref{tab:physpars}. We note that the 3mm continuum is not fitted well by \texttt{Stardust}, which suggests a steep $\beta$ slope or optically thick dust in FIR \citep{Jin2022}.
We thus performed FIR SED fitting with modified blackbody models using the \texttt{Mercurius} code \citep{Witstok2022-mercurius}, accounting for both cases of optically thin and thick dust in FIR \citep{Jin2022}. As shown in Fig.~\ref{fig-thick}, the optically thick model performs slightly better than the thin ones, yielding a dust temperature of $T_{\rm dust}=36.5\pm0.6$~K and a dust mass of $M_{\rm dust}=(2.9\pm0.3)\times10^{9}~M_\odot$. The dust temperature is consistent with the $T_{\rm dust}-z$ relation of main sequence galaxies by \cite{Schreiber2018Tdust}.

To constrain the magnification $\mu$, we performed lens modeling of the F777W image by adopting the methodology from \cite{Vegetti2009lensing} and \cite{Rizzo2018lensing}. As shown in Fig.~\ref{fig-lens}, the lensing arc is well modeled with a magnification factor of $\mu=3.6^{+2.0}_{-0.9}$. This magnification is consistent with the findings of \cite{Pearson2024cosbo7}, which are based on F777W data, but is slightly higher than the value estimated by \cite{Ling2024cosbo7} from the F444W image. 

We derive the molecular gas mass using [CI](2-1) and CO(3-2) as gas tracers: (1) Adopting a $R_{\rm [CI]}=L'_{\rm [CI](2-1)}/L'_{\rm [CI](1-0)}=0.3\pm0.1$ \citep{Jiao2019CI} and assuming the excitation $T_{\rm exc}=T_{\rm dust, thick}$, we obtain a gas mass of $M_{\rm gas,CI}=(5.04\pm1.70)\times10^{11}~M_\odot$ using the scaling relation from \cite{Valentino2018CI}. (2) Assuming a CO line ratio of $r_{\rm 31}=0.84\pm0.26$ from \cite{Riechers2020COLF}, we obtain a CO(1-0) luminosity of $L'_{\rm CO(1-0)}=(7.7\pm2.4)\times10^{11}$~K~km~s$^{-1}$~pc$^2$, which gives a gas mass of $M_{\rm gas,\alpha_{CO}=3.6}=(2.8\pm0.9)\times10^{12}~M_\odot$, or ${ M_{\rm gas,\alpha_{CO}=0.4}=(3.1\pm1.0)\times10^{11}~M_\odot}$. The CO-derived gas masses agree with $M_{\rm gas,CI}$ within the uncertainty of $\alpha_{\rm CO}$. 
We adopted a conservative ${\rm SFR= 521_{-42}^{+122}~M_\odot~yr^{-1}}$ comprising the outputs and uncertainties from both \texttt{Stardust} and \texttt{Mercurius}.
As both $r_{31}$ and $\alpha_{\rm CO}$ are uncertain, we simply adopted $M_{\rm gas,CI}$ and derived a median gas depletion time of $\tau\sim1$~Gyr with a lower limit of ${\tau>326}$~Myr. 
This indicates that COSBO-7 is a gas-rich galaxy with a star formation efficiency (SFE) typical of main sequence galaxies \citep{Sargent2014,Magdis2012SED,Magdis2017}. Accounting for the lensing magnification, the stellar mass and SFR of COSBO-7 are consistent with the main sequence at $z\sim2.6$ \citep{Schreiber2017z4MS}. This again suggests that COSBO-7 is a typical dusty star-forming galaxy at $z\sim2.6$ (e.g., \citealt{daCunha2015}).

\begin{figure*}[ht]%
\centering
\includegraphics[width=0.99\textwidth]{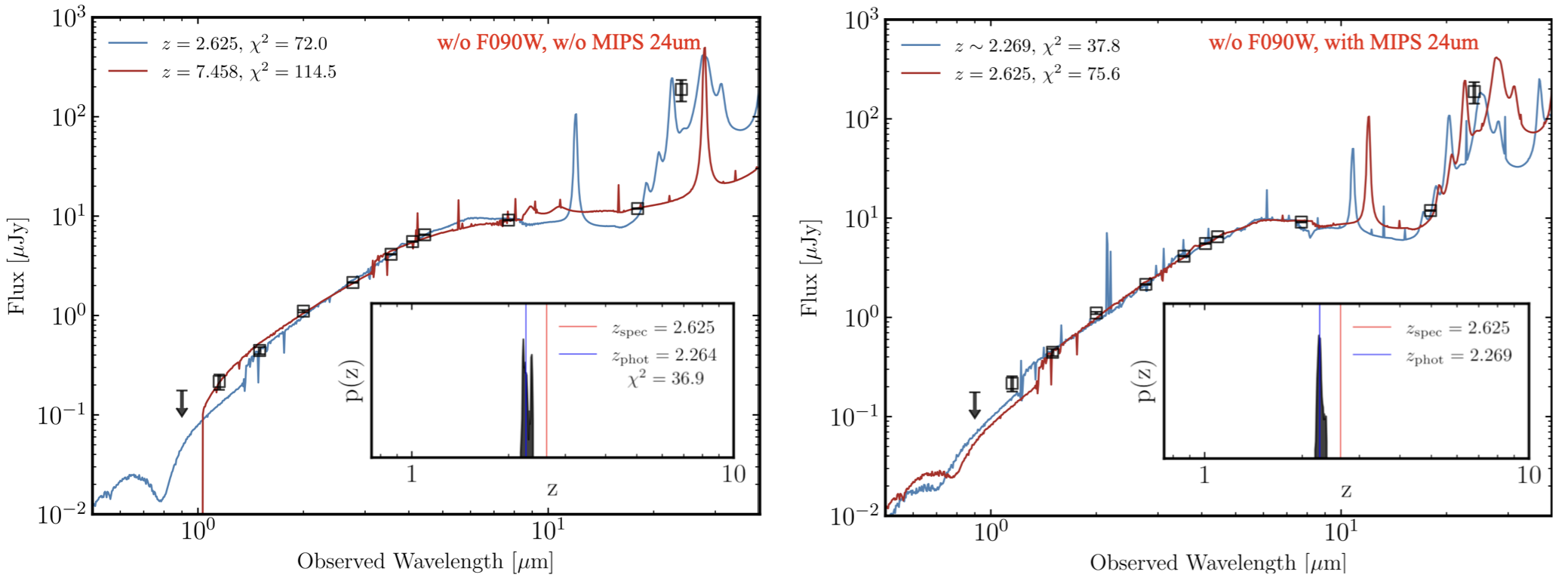}
\caption{NIR-to-MIR SED of COSBO-7 fitted with \texttt{Bagpipes}. {\it Left:} Fitting without the F090W upper limit and without $24\mu$m photometry with the PDF($z$) shown in the subpanel. We also show the SEDs and $\chi^2$ for both $z=2.625$ and $7.458$ cases. {\it Right:} Best fit with MIPS $24\mu$m photometry and without the F090W limit. We present the PDF($z$) with $z_{\rm spec}$ in the subpanel.}
\label{fig:NIR-sed}
\end{figure*}

\section{Discussion}


\subsection{Possible causes of the catastrophic photo-$z$ failure}
\label{discuss1}

The robust $z_{\rm spec}=2.625$ for COSBO-7 derived in this study indicates that the photo-$z\gtrsim7.0$ is a catastrophic failure and highlights that caution should be exercised in studies of DSFGs that rely on photometric redshifts. Here we attempt to uncover the reasons behind the photo-$z$ failure and provide suggestions as to how to see through similar ``cosmic conspiracies''.

Interestingly, we find that the redshift probability distribution function PDF($z$) by \cite{Ling2024cosbo7} indeed shows an insignificant peak at $z\sim2.6$ from \texttt{EAZY} and \texttt{CIGALE} results. As already tested by \cite{Ling2024cosbo7}, the PDF($z$) at $z\sim2.6$ became dominant if adopting the limit of $z<6$ with \texttt{EAZY}, \texttt{CIGALE,} and \texttt{Bagpipes}.
On the other hand, \cite{Ling2024cosbo7} performed SED fitting with two different F090W upper limits, one of which is the average $2\sigma$ depth of the image measured in a $r=0.2''$ aperture, while the other is a $2\sigma$ limit measured within the aperture used for the arc. These authors found that, in either case, the best PDF($z$) solution remains peaked at $z\sim7.0$, and that adjusting the F090W upper limit with a large aperture does not appear to improve the photo-$z$ outputs. We also tested with \texttt{EAZY} by adopting the F090W $3\sigma$ upper limit, and found the best-fit output remains $z\sim7-8$, consistent with \cite{Ling2024cosbo7}. 
As shown in the left panel of Fig.~\ref{fig:NIR-sed}, we tested fitting NIR-to-MIR SEDs at both $z=2.625$ and $z=7.458$, finding that the SED can be fitted at both redshifts, with a subtle difference in the $\chi^2$ values between the two solutions. 
This indicates that the NIR-to-MIR photometry is fully degenerated between $z\sim2.6$ and $z\sim7.5$. 
Further, we test \texttt{Bagpipes} fitting without the F090W upper limit. Interestingly, we find that the PDF($z$) peaks at $z\sim2.3$ without a secondary solution at $z>7$, as in the left panel of Fig.~\ref{fig:NIR-sed}. This photo-$z$ is close to the $z_{\rm spec}=2.625$, and consistent within a typical uncertainty of $\Delta z/(1+z)<10\%$.
This well-recovered photo-$z$ suggests that the F090W flux limit by \cite{Ling2024cosbo7} might be underestimated. We note that the F090W limit in Fig.~\ref{fig:NIR-sed} is a $3\sigma$ upper limit, which is well above the best-fit models of either $z=2.625$ or $z=7.458$, while \cite{Ling2024cosbo7} adopted a stricter $2\sigma$ upper limit. This stringent upper limit likely forced the templates to interpret the data at $\lambda <1\,\mu$m as a Lyman break at $z\sim7$, while excluding solutions at lower redshifts.

Given that the 24$\mu$m flux density is about ten times higher than the F1800W one, we suspect that such an excess, boosted by PAH emission, might be useful to improve the photo-$z$ quality. Therefore, we tested fitting the SED by including the MIPS 24$\mu$m photometry $S_{\rm 24\mu m}=188.4\pm45.9~\mu$Jy measured by \cite{Jin2018cosmos}, which was not used by \cite{Ling2024cosbo7}. Strikingly, this again yields a $z_{\rm phot}=2.3$ (Fig.~\ref{fig:NIR-sed}, right), and the $\chi^2$ is two times smaller than that of the fitting at fixed $z=2.625$ without 24$\mu$m. This is evident in the left panel of Fig.~\ref{fig:NIR-sed}: it is difficult to fit the 24$\mu$m data point with the narrow rest-frame 3.3$\mu$m PAH at $z>7$, while a better fit is achieved with the broad PAH features at rest-frame 6--8$\mu$m. Therefore, the inclusion of $24\mu$m photometry can directly exclude the deceiving $z>7$ solution without fine-tuning the fluxes and flux uncertainties from \cite{Ling2024cosbo7}, and is a straightforward remedy for avoiding the erroneous photo-$z$ of COSBO-7 and could prevent similar situations other objects.
This demonstrates that long-wavelength MIR photometry can significantly improve the photo-$z$ of DSFGs.

It remains puzzling that the best-fit photo-$z\sim2.3$ does not agree well with $z_{\rm spec}=2.625$. In the right panel of Fig.~\ref{fig:NIR-sed}, we show the SED fitting at both redshifts. Interestingly, a significant discrepancy is found for F115W, where the F115W measurement from \cite{Ling2024cosbo7} is clearly above the SED of $z_{\rm spec}=2.625$. As no strong line is expected in the F115W filter, this discrepancy suggests that the F115W flux is overestimated or that the flux uncertainty is underestimated. Given that this F115W photometry was better fitted by $z>7$ models \citep{Ling2024cosbo7}, the F115W measurement is likely problematic, and could also be a reason for the photo-$z$ failure.

\subsection{Caution on photo-$z$}

As COSBO-7 is a strongly lensed galaxy with multiwavelength JWST photometry spanning NIR to MIR, it is alarming that the photo-$z$ failed dramatically, undermining the physical parameters relying on it. Given that COSBO-7 is a typical DSFG at $z\sim2.6$, such photo-$z$ failure could occur in other galaxies. 
As depicted by the PDF($z$) of COSBO-7 from \cite{Ling2024cosbo7}, there are no visible peaks at $z<5$ from \texttt{LePhare} and $z<4$ from \texttt{Bagpipes}; however, the statistically disfavored peak around $z\sim2.6$ from \texttt{EAZY} and \texttt{CIGALE} is now proven to be closer to the real solution. This highlights that caution should be exercised when interpreting the output of photo-$z$ codes, and low-$z$ solutions that appear statistically insignificant in the PDF($z$) cannot be ruled out.

As demonstrated in Sect.~\ref{discuss1}, it is also remarkable that a subtle adjustment of the F090W upper limit can tremendously impact the robustness of the photo-$z$. Finally, it is also clear that a combination of photo-$z$ with a single line detection in the mm might not be sufficient for a robust determination of the redshift of DSFGs, especially if the emission line is consistent with multiple solutions. Indeed, this combination entails the danger of providing a deceiving preference for the most exotic but wrong solution. 
We note that \cite{Ling2024cosbo7} did highlight uncertainties regarding the $z>7$ solution and pointed out that low-z solutions cannot be totally ruled out, although their work only shows that high-z solutions are favored.

It is unclear as to whether COSBO-7 is a rare case or similar catastrophic photo-$z$ failures are common among DSFGs. A large sample is required to determine whether these failures are prevalent or even systematic. If such failures were found to be systematic, literature studies relying purely on photo-$z$ and pseudo-spectroscopic redshifts from single line detection would need to be revised.

\section{Conclusions}

Using ALMA observations, we confirmed that the $z>7$ DSFG candidate COSBO-7 is, in reality, at $z=2.625$. Our conclusions are as follows:

1. We detect three lines with a high level of certainty and identify them as CO(3-2), CO(7-6), and [CI](2-1) at $z=2.625$, thereby robustly confirming the redshift of COSBO-7. This is in tension with the  photometric redshift of $z\gtrsim7.0$ reported by \cite{Ling2024cosbo7}.

2. With the confirmed redshift, we derive physical parameters for COSBO-7 and find it to be a main sequence galaxy with possible optically thick dust.

3. We examined possible explanations for the catastrophic photo-$z$ failure, and attribute it to (1) the likely underestimation of the F090W upper limit and the F115W flux uncertainty; and (2) the lack of photometry at wavelengths beyond 20$\mu$m sampling the PAH features at $z\sim2.6$.

4. Notably, we recover an almost accurate $z_{\rm phot}\sim2.3$ by including the MIPS 24$\mu$m photometry without applying further changes with respect to the literature photometry. This provides a straightforward remedy for the erroneous photo-$z$, and demonstrates the importance of long-wavelength MIR data in supporting photo-$z$ measurements.

This work highlights a common challenge in modeling the SEDs of DSFGs, and provides a cautionary tale regarding the reliability of photometric redshifts and redshifts that rely on single line detections. Long-wavelength MIR photometry can significantly improve the photo-$z$ quality, and so we encourage the use of MIPS or MIRI 24$\mu$m in SED fitting. 
However, even with this additional sampling, the photo-$z$ accuracy is still dependent on the sufficient dominance of certain spectral features.
As such, detecting multiple lines remains the only way to unambiguously identify the redshifts of DSFGs, and the future Wideband Sensitivity Upgrade of ALMA will turn it into an even more powerful "redshift machine".

\begin{acknowledgements}
We thank Haojing Yan, Emanuele Daddi, Jorge Zavala and Ian Smail for helpful discussions in the preparation of this manuscript. We thank Frank Bertoldi for constructive and insightful review of this manuscript.
This paper makes use of the following ALMA data: ADS/JAO.ALMA\#2023.A.00021, 2022.1.00863, 2021.1.00705.S, and 2016.1.00463.S. ALMA is a partnership of ESO (representing its member states), NSF (USA) and NINS (Japan), together with NRC (Canada), NSTC and ASIAA (Taiwan), and KASI (Republic of Korea), in cooperation with the Republic of Chile. The Joint ALMA Observatory is operated by ESO, AUI/NRAO and NAOJ.
SJ acknowledges financial support from the European Union's Horizon Europe research and innovation program under the Marie Sk\l{}odowska-Curie grant No. 101060888.
JH acknowledges support from
the ERC Consolidator Grant 101088676 (VOYAJ).
GEM and SJ acknowledge the Villum Fonden research grants 37440 and 13160.
The Cosmic Dawn Center (DAWN) is funded by the Danish National Research Foundation under grant DNRF140.
APV acknowledges support from the Carlsberg Foundation (grant no CF20-0534).

\end{acknowledgements}

\bibliographystyle{aa}
\bibliography{biblio}

\begin{thebibliography}{74}
\expandafter\ifx\csname natexlab\endcsname\relax\def\natexlab#1{#1}\fi

\bibitem[{{Akins} {et~al.}(2023){Akins}, {Casey}, {Allen}, {Bagley}, {Dickinson}, {Finkelstein}, {Franco}, {Harish}, {Arrabal Haro}, {Ilbert}, {Kartaltepe}, {Koekemoer}, {Liu}, {Long}, {McCracken}, {Paquereau}, {Papovich}, {Pirzkal}, {Rhodes}, {Robertson}, {Shuntov}, {Toft}, {Yang}, {Barro}, {Bisigello}, {Buat}, {Champagne}, {Cooper}, {Costantin}, {de La Vega}, {Drakos}, {Faisst}, {Fontana}, {Fujimoto}, {Gillman}, {G{\'o}mez-Guijarro}, {Gozaliasl}, {Hathi}, {Hayward}, {Hirschmann}, {Holwerda}, {Jin}, {Kocevski}, {Kokorev}, {Lambrides}, {Lucas}, {Magdis}, {Magnelli}, {McKinney}, {Mobasher}, {P{\'e}rez-Gonz{\'a}lez}, {Rich}, {Seill{\'e}}, {Talia}, {Urry}, {Valentino}, {Whitaker}, {Yung}, {Zavala}, {Cosmos-Web Team}, \& {Ceers Team}}]{Akins2023z8candi}
{Akins}, H.~B., {Casey}, C.~M., {Allen}, N., {et~al.} 2023, \apj, 956, 61

\bibitem[{{Aretxaga} {et~al.}(2011){Aretxaga}, {Wilson}, {Aguilar}, {Alberts}, {Scott}, {Scoville}, {Yun}, {Austermann}, {Downes}, {Ezawa}, {Hatsukade}, {Hughes}, {Kawabe}, {Kohno}, {Oshima}, {Perera}, {Tamura}, \& {Zeballos}}]{Aretxaga2011}
{Aretxaga}, I., {Wilson}, G.~W., {Aguilar}, E., {et~al.} 2011, \mnras, 415, 3831

\bibitem[{{Arnouts} {et~al.}(1999){Arnouts}, {Cristiani}, {Moscardini}, {Matarrese}, {Lucchin}, {Fontana}, \& {Giallongo}}]{Arnouts1999LePhare}
{Arnouts}, S., {Cristiani}, S., {Moscardini}, L., {et~al.} 1999, \mnras, 310, 540

\bibitem[{{Arrabal Haro} {et~al.}(2023){Arrabal Haro}, {Dickinson}, {Finkelstein}, {Kartaltepe}, {Donnan}, {Burgarella}, {Carnall}, {Cullen}, {Dunlop}, {Fern{\'a}ndez}, {Fujimoto}, {Jung}, {Krips}, {Larson}, {Papovich}, {P{\'e}rez-Gonz{\'a}lez}, {Amor{\'\i}n}, {Bagley}, {Buat}, {Casey}, {Chworowsky}, {Cohen}, {Ferguson}, {Giavalisco}, {Huertas-Company}, {Hutchison}, {Kocevski}, {Koekemoer}, {Lucas}, {McLeod}, {McLure}, {Pirzkal}, {Seill{\'e}}, {Trump}, {Weiner}, {Wilkins}, \& {Zavala}}]{Arrabal_Haro2023Natur}
{Arrabal Haro}, P., {Dickinson}, M., {Finkelstein}, S.~L., {et~al.} 2023, \nat, 622, 707

\bibitem[{{Atek} {et~al.}(2023){Atek}, {Shuntov}, {Furtak}, {Richard}, {Kneib}, {Mahler}, {Zitrin}, {McCracken}, {Charlot}, {Chevallard}, \& {Chemerynska}}]{Atek2023z16}
{Atek}, H., {Shuntov}, M., {Furtak}, L.~J., {et~al.} 2023, \mnras, 519, 1201

\bibitem[{{Barrufet} {et~al.}(2023){Barrufet}, {Oesch}, {Weibel}, {Brammer}, {Bezanson}, {Bouwens}, {Fudamoto}, {Gonzalez}, {Gottumukkala}, {Illingworth}, {Heintz}, {Holden}, {Labbe}, {Magee}, {Naidu}, {Nelson}, {Stefanon}, {Smit}, {van Dokkum}, {Weaver}, \& {Williams}}]{Barrufet2023dark}
{Barrufet}, L., {Oesch}, P.~A., {Weibel}, A., {et~al.} 2023, \mnras, 522, 449

\bibitem[{{Bertoldi} {et~al.}(2007){Bertoldi}, {Carilli}, {Aravena}, {Schinnerer}, {Voss}, {Smolcic}, {Jahnke}, {Scoville}, {Blain}, {Menten}, {Lutz}, {Brusa}, {Taniguchi}, {Capak}, {Mobasher}, {Lilly}, {Thompson}, {Aussel}, {Kreysa}, {Hasinger}, {Aguirre}, {Schlaerth}, \& {Koekemoer}}]{Bertoldi2007}
{Bertoldi}, F., {Carilli}, C., {Aravena}, M., {et~al.} 2007, \apjs, 172, 132

\bibitem[{{Boone} {et~al.}(2011){Boone}, {Schaerer}, {Pell{\'o}}, {Lutz}, {Weiss}, {Egami}, {Smail}, {Rex}, {Rawle}, {Ivison}, {Laporte}, {Beelen}, {Combes}, {Blain}, {Richard}, {Kneib}, {Zamojski}, {Dessauges-Zavadsky}, {Altieri}, {van der Werf}, {Swinbank}, {P{\'e}rez-Gonz{\'a}lez}, {Clement}, {Nordon}, {Magnelli}, \& {Menten}}]{Boone2011}
{Boone}, F., {Schaerer}, D., {Pell{\'o}}, R., {et~al.} 2011, \aap, 534, A124

\bibitem[{{Boquien} {et~al.}(2019){Boquien}, {Burgarella}, {Roehlly}, {Buat}, {Ciesla}, {Corre}, {Inoue}, \& {Salas}}]{Boquien2019CIGALE}
{Boquien}, M., {Burgarella}, D., {Roehlly}, Y., {et~al.} 2019, \aap, 622, A103

\bibitem[{{Brammer} {et~al.}(2008){Brammer}, {van Dokkum}, \& {Coppi}}]{Brammer2008EAZY}
{Brammer}, G.~B., {van Dokkum}, P.~G., \& {Coppi}, P. 2008, \apj, 686, 1503

\bibitem[{{Carnall} {et~al.}(2018){Carnall}, {McLure}, {Dunlop}, \& {Dav{\'e}}}]{Carnall2018Bagpipes}
{Carnall}, A.~C., {McLure}, R.~J., {Dunlop}, J.~S., \& {Dav{\'e}}, R. 2018, \mnras, 480, 4379

\bibitem[{{Carniani} {et~al.}(2024){Carniani}, {Hainline}, {D'Eugenio}, {Eisenstein}, {Jakobsen}, {Witstok}, {Johnson}, {Chevallard}, {Maiolino}, {Helton}, {Willott}, {Robertson}, {Alberts}, {Arribas}, {Baker}, {Bhatawdekar}, {Boyett}, {Bunker}, {Cameron}, {Cargile}, {Charlot}, {Curti}, {Curtis-Lake}, {Egami}, {Giardino}, {Isaak}, {Ji}, {Jones}, {Kumari}, {Maseda}, {Parlanti}, {P{\'e}rez-Gonz{\'a}lez}, {Rawle}, {Rieke}, {Rieke}, {Del Pino}, {Saxena}, {Scholtz}, {Smit}, {Sun}, {Tacchella}, {{\"U}bler}, {Venturi}, {Williams}, \& {Willmer}}]{Carniani2024z14}
{Carniani}, S., {Hainline}, K., {D'Eugenio}, F., {et~al.} 2024, \nat, 633, 318

\bibitem[{{Casey} {et~al.}(2024){Casey}, {Akins}, {Shuntov}, {Ilbert}, {Paquereau}, {Franco}, {Hayward}, {Finkelstein}, {Boylan-Kolchin}, {Robertson}, {Allen}, {Brinch}, {Cooper}, {Ding}, {Drakos}, {Faisst}, {Fujimoto}, {Gillman}, {Harish}, {Hirschmann}, {Jin}, {Kartaltepe}, {Koekemoer}, {Kokorev}, {Liu}, {Long}, {Magdis}, {Maraston}, {Martin}, {McCracken}, {McKinney}, {Mobasher}, {Rhodes}, {Rich}, {Sanders}, {Silverman}, {Toft}, {Vijayan}, {Weaver}, {Wilkins}, {Yang}, \& {Zavala}}]{Casey2024z10}
{Casey}, C.~M., {Akins}, H.~B., {Shuntov}, M., {et~al.} 2024, \apj, 965, 98

\bibitem[{{Casey} {et~al.}(2023){Casey}, {Kartaltepe}, {Drakos}, {Franco}, {Harish}, {Paquereau}, {Ilbert}, {Rose}, {Cox}, {Nightingale}, {Robertson}, {Silverman}, {Koekemoer}, {Massey}, {McCracken}, {Rhodes}, {Akins}, {Allen}, {Amvrosiadis}, {Arango-Toro}, {Bagley}, {Bongiorno}, {Capak}, {Champagne}, {Chartab}, {Ch{\'a}vez Ortiz}, {Chworowsky}, {Cooke}, {Cooper}, {Darvish}, {Ding}, {Faisst}, {Finkelstein}, {Fujimoto}, {Gentile}, {Gillman}, {Gould}, {Gozaliasl}, {Hayward}, {He}, {Hemmati}, {Hirschmann}, {Jahnke}, {Jin}, {Khostovan}, {Kokorev}, {Lambrides}, {Laigle}, {Larson}, {Leung}, {Liu}, {Liaudat}, {Long}, {Magdis}, {Mahler}, {Mainieri}, {Manning}, {Maraston}, {Martin}, {McCleary}, {McKinney}, {McPartland}, {Mobasher}, {Pattnaik}, {Renzini}, {Rich}, {Sanders}, {Sattari}, {Scognamiglio}, {Scoville}, {Sheth}, {Shuntov}, {Sparre}, {Suzuki}, {Talia}, {Toft}, {Trakhtenbrot}, {Urry}, {Valentino}, {Vanderhoof}, {Vardoulaki}, {Weaver}, {Whitaker}, {Wilkins}, {Yang}, \& {Zavala}}]{Casey2023cosmosweb}
{Casey}, C.~M., {Kartaltepe}, J.~S., {Drakos}, N.~E., {et~al.} 2023, \apj, 954, 31

\bibitem[{{Chabrier}(2003)}]{Chabrier2003}
{Chabrier}, G. 2003, \pasp, 115, 763

\bibitem[{{Chakraborty} {et~al.}(2024){Chakraborty}, {Sarkar}, {Wolk}, {Schneider}, {Brickhouse}, {Lanzetta}, {Foster}, \& {Smith}}]{Chakraborty2024z11}
{Chakraborty}, P., {Sarkar}, A., {Wolk}, S., {et~al.} 2024, arXiv e-prints, arXiv:2406.05306

\bibitem[{{Chen} {et~al.}(2022){Chen}, {Liao}, {Smail}, {Swinbank}, {Ao}, {Bunker}, {Chapman}, {Hatsukade}, {Ivison}, {Lee}, {Serjeant}, {Umehata}, {Wang}, \& {Zhao}}]{ChenCC2022alma}
{Chen}, C.-C., {Liao}, C.-L., {Smail}, I., {et~al.} 2022, \apj, 929, 159

\bibitem[{{Cox} {et~al.}(2023){Cox}, {Neri}, {Berta}, {Ismail}, {Stanley}, {Young}, {Jin}, {Bakx}, {Beelen}, {Dannerbauer}, {Krips}, {Lehnert}, {Omont}, {Riechers}, {Baker}, {Bendo}, {Borsato}, {Buat}, {Butler}, {Chartab}, {Cooray}, {Dye}, {Eales}, {Gavazzi}, {Hughes}, {Ivison}, {Jones}, {Marchetti}, {Messias}, {Nanni}, {Negrello}, {Perez-Fournon}, {Serjeant}, {Urquhart}, {Vlahakis}, {Wei{\ss}}, {van der Werf}, \& {Yang}}]{Cox2023zGal}
{Cox}, P., {Neri}, R., {Berta}, S., {et~al.} 2023, \aap, 678, A26

\bibitem[{{da Cunha} {et~al.}(2008){da Cunha}, {Charlot}, \& {Elbaz}}]{da_Cunha2008magphys}
{da Cunha}, E., {Charlot}, S., \& {Elbaz}, D. 2008, \mnras, 388, 1595

\bibitem[{{da Cunha} {et~al.}(2015){da Cunha}, {Walter}, {Smail}, {Swinbank}, {Simpson}, {Decarli}, {Hodge}, {Weiss}, {van der Werf}, {Bertoldi}, {Chapman}, {Cox}, {Danielson}, {Dannerbauer}, {Greve}, {Ivison}, {Karim}, \& {Thomson}}]{daCunha2015}
{da Cunha}, E., {Walter}, F., {Smail}, I.~R., {et~al.} 2015, \apj, 806, 110

\bibitem[{{Delvecchio} {et~al.}(2021){Delvecchio}, {Daddi}, {Sargent}, {Jarvis}, {Elbaz}, {Jin}, {Liu}, {Whittam}, {Algera}, {Carraro}, {D'Eugenio}, {Delhaize}, {Kalita}, {Leslie}, {Moln{\'a}r}, {Novak}, {Prandoni}, {Smol{\v{c}}i{\'c}}, {Ao}, {Aravena}, {Bournaud}, {Collier}, {Randriamampandry}, {Randriamanakoto}, {Rodighiero}, {Schober}, {White}, \& {Zamorani}}]{Delvecchio2021}
{Delvecchio}, I., {Daddi}, E., {Sargent}, M.~T., {et~al.} 2021, \aap, 647, A123

\bibitem[{{Donnan} {et~al.}(2023){Donnan}, {McLeod}, {Dunlop}, {McLure}, {Carnall}, {Begley}, {Cullen}, {Hamadouche}, {Bowler}, {Magee}, {McCracken}, {Milvang-Jensen}, {Moneti}, \& {Targett}}]{Donnan2023z16}
{Donnan}, C.~T., {McLeod}, D.~J., {Dunlop}, J.~S., {et~al.} 2023, \mnras, 518, 6011

\bibitem[{{Dudzevi{\v{c}}i{\={u}}t{\.{e}}} {et~al.}(2020){Dudzevi{\v{c}}i{\={u}}t{\.{e}}}, {Smail}, {Swinbank}, {Stach}, {Almaini}, {da Cunha}, {An}, {Arumugam}, {Birkin}, {Blain}, {Chapman}, {Chen}, {Conselice}, {Coppin}, {Dunlop}, {Farrah}, {Geach}, {Gullberg}, {Hartley}, {Hodge}, {Ivison}, {Maltby}, {Scott}, {Simpson}, {Simpson}, {Thomson}, {Walter}, {Wardlow}, {Weiss}, \& {van der Werf}}]{Dudzeviciute2020SMG}
{Dudzevi{\v{c}}i{\={u}}t{\.{e}}}, U., {Smail}, I., {Swinbank}, A.~M., {et~al.} 2020, \mnras, 494, 3828

\bibitem[{{Dunlop} {et~al.}(2021){Dunlop}, {Abraham}, {Ashby}, {Bagley}, {Best}, {Bongiorno}, {Bouwens}, {Bowler}, {Brammer}, {Bremer}, {Calabro'}, {Carnall}, {Castellano}, {Cirasuolo}, {Conselice}, {Cullen}, {Dave}, {Dayal}, {Dekel}, {Dickinson}, {Duncan}, {Elbaz}, {Ellis}, {Ferguson}, {Ferrara}, {Finkelstein}, {Fontana}, {Furlanetto}, {Fynbo}, {Gallerani}, {Gardner}, {Giavalisco}, {Grazian}, {Grogin}, {Harikane}, {Hopkins}, {Ilbert}, {Illingworth}, {Juneau}, {Jung}, {Kartaltepe}, {Kassin}, {Kauffmann}, {Khochfar}, {Kirkpatrick}, {Kocevski}, {Koekemoer}, {Labbe}, {Laporte}, {Larson}, {Lucas}, {Magee}, {Mason}, {McCracken}, {McLeod}, {McLure}, {Merlin}, {Mesinger}, {Milvang-Jensen}, {Newman}, {Oesch}, {Ouchi}, {Pacifici}, {Papovich}, {Peacock}, {Peeples}, {Pentericci}, {Perez-Gonzalez}, {Pirzkal}, {Pope}, {Pye}, {Reddy}, {Robertson}, {Salvato}, {Santini}, {Schaerer}, {Shapley}, {Simons}, {Smit}, {Smith}, {Snyder}, {Somerville}, {Stanway}, {Stefanon}, {Tasca}, {Tikkanen}, {Tresse}, {Trump}, {Whitaker},
  {Wilkins}, {Wright}, {Wyithe}, {van Dokkum}, \& {van der Werf}}]{Dunlop2021PRIMER}
{Dunlop}, J.~S., {Abraham}, R.~G., {Ashby}, M. L.~N., {et~al.} 2021, {PRIMER: Public Release IMaging for Extragalactic Research}, JWST Proposal. Cycle 1, ID. \#1837

\bibitem[{{Endsley} {et~al.}(2022){Endsley}, {Stark}, {Fan}, {Smit}, {Wang}, {Yang}, {Hainline}, {Lyu}, {Bouwens}, \& {Schouws}}]{Endsley2022z68}
{Endsley}, R., {Stark}, D.~P., {Fan}, X., {et~al.} 2022, \mnras, 512, 4248

\bibitem[{{Finkelstein} {et~al.}(2022){Finkelstein}, {Bagley}, {Arrabal Haro}, {Dickinson}, {Ferguson}, {Kartaltepe}, {Papovich}, {Burgarella}, {Kocevski}, {Huertas-Company}, {Iyer}, {Koekemoer}, {Larson}, {P{\'e}rez-Gonz{\'a}lez}, {Rose}, {Tacchella}, {Wilkins}, {Chworowsky}, {Medrano}, {Morales}, {Somerville}, {Yung}, {Fontana}, {Giavalisco}, {Grazian}, {Grogin}, {Kewley}, {Kirkpatrick}, {Kurczynski}, {Lotz}, {Pentericci}, {Pirzkal}, {Ravindranath}, {Ryan}, {Trump}, {Yang}, {Almaini}, {Amor{\'\i}n}, {Annunziatella}, {Backhaus}, {Barro}, {Behroozi}, {Bell}, {Bhatawdekar}, {Bisigello}, {Bromm}, {Buat}, {Buitrago}, {Calabr{\`o}}, {Casey}, {Castellano}, {Ch{\'a}vez Ortiz}, {Ciesla}, {Cleri}, {Cohen}, {Cole}, {Cooke}, {Cooper}, {Cooray}, {Costantin}, {Cox}, {Croton}, {Daddi}, {Dav{\'e}}, {de La Vega}, {Dekel}, {Elbaz}, {Estrada-Carpenter}, {Faber}, {Fern{\'a}ndez}, {Finkelstein}, {Freundlich}, {Fujimoto}, {Garc{\'\i}a-Argum{\'a}nez}, {Gardner}, {Gawiser}, {G{\'o}mez-Guijarro}, {Guo}, {Hamblin}, {Hamilton},
  {Hathi}, {Holwerda}, {Hirschmann}, {Hutchison}, {Jaskot}, {Jha}, {Jogee}, {Juneau}, {Jung}, {Kassin}, {Le Bail}, {Leung}, {Lucas}, {Magnelli}, {Mantha}, {Matharu}, {McGrath}, {McIntosh}, {Merlin}, {Mobasher}, {Newman}, {Nicholls}, {Pandya}, {Rafelski}, {Ronayne}, {Santini}, {Seill{\'e}}, {Shah}, {Shen}, {Simons}, {Snyder}, {Stanway}, {Straughn}, {Teplitz}, {Vanderhoof}, {Vega-Ferrero}, {Wang}, {Weiner}, {Willmer}, {Wuyts}, {Zavala}, \& {Ceers Team}}]{Finkelstein2022z12}
{Finkelstein}, S.~L., {Bagley}, M.~B., {Arrabal Haro}, P., {et~al.} 2022, \apjl, 940, L55

\bibitem[{{Fudamoto} {et~al.}(2021){Fudamoto}, {Oesch}, {Schouws}, {Stefanon}, {Smit}, {Bouwens}, {Bowler}, {Endsley}, {Gonzalez}, {Inami}, {Labbe}, {Stark}, {Aravena}, {Barrufet}, {da Cunha}, {Dayal}, {Ferrara}, {Graziani}, {Hodge}, {Hutter}, {Li}, {De Looze}, {Nanayakkara}, {Pallottini}, {Riechers}, {Schneider}, {Ucci}, {van der Werf}, \& {White}}]{Fudamoto2021Nature}
{Fudamoto}, Y., {Oesch}, P.~A., {Schouws}, S., {et~al.} 2021, \nat, 597, 489

\bibitem[{{Geach} {et~al.}(2017){Geach}, {Dunlop}, {Halpern}, {Smail}, {van der Werf}, {Alexander}, {Almaini}, {Aretxaga}, {Arumugam}, {Asboth}, {Banerji}, {Beanlands}, {Best}, {Blain}, {Birkinshaw}, {Chapin}, {Chapman}, {Chen}, {Chrysostomou}, {Clarke}, {Clements}, {Conselice}, {Coppin}, {Cowley}, {Danielson}, {Eales}, {Edge}, {Farrah}, {Gibb}, {Harrison}, {Hine}, {Hughes}, {Ivison}, {Jarvis}, {Jenness}, {Jones}, {Karim}, {Koprowski}, {Knudsen}, {Lacey}, {Mackenzie}, {Marsden}, {McAlpine}, {McMahon}, {Meijerink}, {Micha{\l}owski}, {Oliver}, {Page}, {Peacock}, {Rigopoulou}, {Robson}, {Roseboom}, {Rotermund}, {Scott}, {Serjeant}, {Simpson}, {Simpson}, {Smith}, {Spaans}, {Stanley}, {Stevens}, {Swinbank}, {Targett}, {Thomson}, {Valiante}, {Wake}, {Webb}, {Willott}, {Zavala}, \& {Zemcov}}]{Geach2017scuba2}
{Geach}, J.~E., {Dunlop}, J.~S., {Halpern}, M., {et~al.} 2017, \mnras, 465, 1789

\bibitem[{{G{\'o}mez-Guijarro} {et~al.}(2019){G{\'o}mez-Guijarro}, {Riechers}, {Pavesi}, {Magdis}, {Leung}, {Valentino}, {Toft}, {Aravena}, {Chapman}, {Clements}, {Dannerbauer}, {Oliver}, {P{\'e}rez-Fournon}, \& {Valtchanov}}]{Gomez-Guijarro2019}
{G{\'o}mez-Guijarro}, C., {Riechers}, D.~A., {Pavesi}, R., {et~al.} 2019, \apj, 872, 117

\bibitem[{{Harikane} {et~al.}(2023){Harikane}, {Ouchi}, {Oguri}, {Ono}, {Nakajima}, {Isobe}, {Umeda}, {Mawatari}, \& {Zhang}}]{Harikane2023z10}
{Harikane}, Y., {Ouchi}, M., {Oguri}, M., {et~al.} 2023, \apjs, 265, 5

\bibitem[{{Hygate} {et~al.}(2023){Hygate}, {Hodge}, {da Cunha}, {Rybak}, {Schouws}, {Inami}, {Stefanon}, {Graziani}, {Schneider}, {Dayal}, {Bouwens}, {Smit}, {Bowler}, {Endsley}, {Gonzalez}, {Oesch}, {Stark}, {Algera}, {Aravena}, {Barrufet}, {Ferrara}, {Fudamoto}, {Hilhorst}, {De Looze}, {Nanayakkara}, {Pallottini}, {Riechers}, {Sommovigo}, {Topping}, \& {van der Werf}}]{Hygate2023z7}
{Hygate}, A.~P.~S., {Hodge}, J.~A., {da Cunha}, E., {et~al.} 2023, \mnras, 524, 1775

\bibitem[{{Ilbert} {et~al.}(2006){Ilbert}, {Arnouts}, {McCracken}, {Bolzonella}, {Bertin}, {Le F{\`e}vre}, {Mellier}, {Zamorani}, {Pell{\`o}}, {Iovino}, {Tresse}, {Le Brun}, {Bottini}, {Garilli}, {Maccagni}, {Picat}, {Scaramella}, {Scodeggio}, {Vettolani}, {Zanichelli}, {Adami}, {Bardelli}, {Cappi}, {Charlot}, {Ciliegi}, {Contini}, {Cucciati}, {Foucaud}, {Franzetti}, {Gavignaud}, {Guzzo}, {Marano}, {Marinoni}, {Mazure}, {Meneux}, {Merighi}, {Paltani}, {Pollo}, {Pozzetti}, {Radovich}, {Zucca}, {Bondi}, {Bongiorno}, {Busarello}, {de La Torre}, {Gregorini}, {Lamareille}, {Mathez}, {Merluzzi}, {Ripepi}, {Rizzo}, \& {Vergani}}]{Ilbert2006LePhare}
{Ilbert}, O., {Arnouts}, S., {McCracken}, H.~J., {et~al.} 2006, \aap, 457, 841

\bibitem[{{Jiao} {et~al.}(2019){Jiao}, {Zhao}, {Lu}, {Gao}, {Salak}, {Zhu}, {Zhang}, {Jiang}, \& {Tan}}]{Jiao2019CI}
{Jiao}, Q., {Zhao}, Y., {Lu}, N., {et~al.} 2019, \apj, 880, 133

\bibitem[{{Jin} {et~al.}(2018){Jin}, {Daddi}, {Liu}, {Smol{\v c}i{\'c}}, {Schinnerer}, {Calabr{\`o}}, {Gu}, {Delhaize}, {Delvecchio}, {Gao}, {Salvato}, {Puglisi}, {Dickinson}, {Bertoldi}, {Sargent}, {Novak}, {Magdis}, {Aretxaga}, {Wilson}, \& {Capak}}]{Jin2018cosmos}
{Jin}, S., {Daddi}, E., {Liu}, D., {et~al.} 2018, \apj, 864, 56

\bibitem[{{Jin} {et~al.}(2019){Jin}, {Daddi}, {Magdis}, {Liu}, {Schinnerer}, {Papadopoulos}, {Gu}, {Gao}, \& {Calabr{\`o}}}]{Jin2019alma}
{Jin}, S., {Daddi}, E., {Magdis}, G.~E., {et~al.} 2019, \apj, 887, 144

\bibitem[{{Jin} {et~al.}(2022){Jin}, {Daddi}, {Magdis}, {Liu}, {Weaver}, {Tan}, {Valentino}, {Gao}, {Schinnerer}, {Calabr{\`o}}, {Gu}, \& {Sese}}]{Jin2022}
{Jin}, S., {Daddi}, E., {Magdis}, G.~E., {et~al.} 2022, \aap, 665, A3

\bibitem[{{Kokorev} {et~al.}(2021){Kokorev}, {Magdis}, {Davidzon}, {Brammer}, {Valentino}, {Daddi}, {Ciesla}, {Liu}, {Jin}, {Cortzen}, {Delvecchio}, {Gim{\'e}nez-Arteaga}, {G{\'o}mez-Guijarro}, {Sargent}, {Toft}, \& {Weaver}}]{Kokorev2021stardust}
{Kokorev}, V.~I., {Magdis}, G.~E., {Davidzon}, I., {et~al.} 2021, \apj, 921, 40

\bibitem[{{Kriek} {et~al.}(2009){Kriek}, {van Dokkum}, {Labb{\'e}}, {Franx}, {Illingworth}, {Marchesini}, \& {Quadri}}]{Kriek2009FAST}
{Kriek}, M., {van Dokkum}, P.~G., {Labb{\'e}}, I., {et~al.} 2009, \apj, 700, 221

\bibitem[{{Ling} {et~al.}(2024){Ling}, {Sun}, {Cheng}, {Li}, {Ma}, \& {Yan}}]{Ling2024cosbo7}
{Ling}, C., {Sun}, B., {Cheng}, C., {et~al.} 2024, \apjl, 969, L28

\bibitem[{{Liu} {et~al.}(2019){Liu}, {Lang}, {Magnelli}, {Schinnerer}, {Leslie}, {Fudamoto}, {Bondi}, {Groves}, {Jim{\'e}nez-Andrade}, {Harrington}, {Karim}, {Oesch}, {Sargent}, {Vardoulaki}, {B{\v{a}}descu}, {Moser}, {Bertoldi}, {Battisti}, {da Cunha}, {Zavala}, {Vaccari}, {Davidzon}, {Riechers}, \& {Aravena}}]{Liu2019A3COSMOS}
{Liu}, D., {Lang}, P., {Magnelli}, B., {et~al.} 2019, \apjs, 244, 40

\bibitem[{{Magdis} {et~al.}(2012){Magdis}, {Daddi}, {B{\'e}thermin}, {Sargent}, {Elbaz}, {Pannella}, {Dickinson}, {Dannerbauer}, {da Cunha}, {Walter}, {Rigopoulou}, {Charmandaris}, {Hwang}, \& {Kartaltepe}}]{Magdis2012SED}
{Magdis}, G.~E., {Daddi}, E., {B{\'e}thermin}, M., {et~al.} 2012, \apj, 760, 6

\bibitem[{{Magdis} {et~al.}(2017){Magdis}, {Rigopoulou}, {Daddi}, {Bethermin}, {Feruglio}, {Sargent}, {Dannerbauer}, {Dickinson}, {Elbaz}, {Gomez Guijarro}, {Huang}, {Toft}, \& {Valentino}}]{Magdis2017}
{Magdis}, G.~E., {Rigopoulou}, D., {Daddi}, E., {et~al.} 2017, \aap, 603, A93

\bibitem[{{Marrone} {et~al.}(2018){Marrone}, {Spilker}, {Hayward}, {Vieira}, {Aravena}, {Ashby}, {Bayliss}, {B{\'e}thermin}, {Brodwin}, {Bothwell}, {Carlstrom}, {Chapman}, {Chen}, {Crawford}, {Cunningham}, {De Breuck}, {Fassnacht}, {Gonzalez}, {Greve}, {Hezaveh}, {Lacaille}, {Litke}, {Lower}, {Ma}, {Malkan}, {Miller}, {Morningstar}, {Murphy}, {Narayanan}, {Phadke}, {Rotermund}, {Sreevani}, {Stalder}, {Stark}, {Strandet}, {Tang}, \& {Wei{\ss}}}]{Marrone2017Nature}
{Marrone}, D.~P., {Spilker}, J.~S., {Hayward}, C.~C., {et~al.} 2018, \nat, 553, 51

\bibitem[{{McMullin} {et~al.}(2007){McMullin}, {Waters}, {Schiebel}, {Young}, \& {Golap}}]{McMullin2007CASA}
{McMullin}, J.~P., {Waters}, B., {Schiebel}, D., {Young}, W., \& {Golap}, K. 2007, in Astronomical Society of the Pacific Conference Series, Vol. 376, Astronomical Data Analysis Software and Systems XVI, ed. R.~A. {Shaw}, F.~{Hill}, \& D.~J. {Bell}, 127

\bibitem[{{Miettinen} {et~al.}(2015){Miettinen}, {Smol{\v{c}}i{\'c}}, {Novak}, {Aravena}, {Karim}, {Masters}, {Riechers}, {Bussmann}, {McCracken}, {Ilbert}, {Bertoldi}, {Capak}, {Feruglio}, {Halliday}, {Kartaltepe}, {Navarrete}, {Salvato}, {Sanders}, {Schinnerer}, \& {Sheth}}]{Miettinen2015}
{Miettinen}, O., {Smol{\v{c}}i{\'c}}, V., {Novak}, M., {et~al.} 2015, \aap, 577, A29

\bibitem[{{Naidu} {et~al.}(2022){Naidu}, {Oesch}, {Setton}, {Matthee}, {Conroy}, {Johnson}, {Weaver}, {Bouwens}, {Brammer}, {Dayal}, {Illingworth}, {Barrufet}, {Belli}, {Bezanson}, {Bose}, {Heintz}, {Leja}, {Leonova}, {Marques-Chaves}, {Stefanon}, {Toft}, {van der Wel}, {van Dokkum}, {Weibel}, \& {Whitaker}}]{Naidu2022photoz}
{Naidu}, R.~P., {Oesch}, P.~A., {Setton}, D.~J., {et~al.} 2022, arXiv e-prints, arXiv:2208.02794

\bibitem[{{Neri} {et~al.}(2020){Neri}, {Cox}, {Omont}, {Beelen}, {Berta}, {Bakx}, {Lehnert}, {Baker}, {Buat}, {Cooray}, {Dannerbauer}, {Dunne}, {Dye}, {Eales}, {Gavazzi}, {Harris}, {Herrera}, {Hughes}, {Ivison}, {Jin}, {Krips}, {Lagache}, {Marchetti}, {Messias}, {Negrello}, {Perez-Fournon}, {Riechers}, {Serjeant}, {Urquhart}, {Vlahakis}, {Wei{\ss}}, {van der Werf}, {Yang}, \& {Young}}]{Neri2020noema}
{Neri}, R., {Cox}, P., {Omont}, A., {et~al.} 2020, \aap, 635, A7

\bibitem[{{Pearson} {et~al.}(2024){Pearson}, {Serjeant}, {Wang}, {Gao}, {Babul}, {Chapman}, {Chen}, {Clements}, {Conselice}, {Dunlop}, {Fan}, {Ho}, {Hwang}, {Koprowski}, {Micha{\l}owski}, \& {Shim}}]{Pearson2024cosbo7}
{Pearson}, J., {Serjeant}, S., {Wang}, W.-H., {et~al.} 2024, \mnras, 527, 12044

\bibitem[{{Riechers} {et~al.}(2020){Riechers}, {Boogaard}, {Decarli}, {Gonz{\'a}lez-L{\'o}pez}, {Smail}, {Walter}, {Aravena}, {Carilli}, {Cortes}, {Cox}, {D{\'\i}az-Santos}, {Hodge}, {Inami}, {Ivison}, {Kaasinen}, {Wagg}, {Wei{\ss}}, \& {van der Werf}}]{Riechers2020COLF}
{Riechers}, D.~A., {Boogaard}, L.~A., {Decarli}, R., {et~al.} 2020, \apjl, 896, L21

\bibitem[{{Riechers} {et~al.}(2013){Riechers}, {Bradford}, {Clements}, {Dowell}, {P{\'e}rez-Fournon}, {Ivison}, {Bridge}, {Conley}, {Fu}, {Vieira}, {Wardlow}, {Calanog}, {Cooray}, {Hurley}, {Neri}, {Kamenetzky}, {Aguirre}, {Altieri}, {Arumugam}, {Benford}, {B{\'e}thermin}, {Bock}, {Burgarella}, {Cabrera-Lavers}, {Chapman}, {Cox}, {Dunlop}, {Earle}, {Farrah}, {Ferrero}, {Franceschini}, {Gavazzi}, {Glenn}, {Solares}, {Gurwell}, {Halpern}, {Hatziminaoglou}, {Hyde}, {Ibar}, {Kov{\'a}cs}, {Krips}, {Lupu}, {Maloney}, {Martinez-Navajas}, {Matsuhara}, {Murphy}, {Naylor}, {Nguyen}, {Oliver}, {Omont}, {Page}, {Petitpas}, {Rangwala}, {Roseboom}, {Scott}, {Smith}, {Staguhn}, {Streblyanska}, {Thomson}, {Valtchanov}, {Viero}, {Wang}, {Zemcov}, \& {Zmuidzinas}}]{Riechers2013Nature}
{Riechers}, D.~A., {Bradford}, C.~M., {Clements}, D.~L., {et~al.} 2013, \nat, 496, 329

\bibitem[{{Riechers} {et~al.}(2017){Riechers}, {Leung}, {Ivison}, {P{\'e}rez-Fournon}, {Lewis}, {Marques-Chaves}, {Oteo}, {Clements}, {Cooray}, {Greenslade}, {Mart{\'{\i}}nez-Navajas}, {Oliver}, {Rigopoulou}, {Scott}, \& {Weiss}}]{Riechers2017}
{Riechers}, D.~A., {Leung}, T.~K.~D., {Ivison}, R.~J., {et~al.} 2017, \apj, 850, 1

\bibitem[{{Rizzo} {et~al.}(2018){Rizzo}, {Vegetti}, {Fraternali}, \& {Di Teodoro}}]{Rizzo2018lensing}
{Rizzo}, F., {Vegetti}, S., {Fraternali}, F., \& {Di Teodoro}, E. 2018, \mnras, 481, 5606

\bibitem[{{Rowland} {et~al.}(2024){Rowland}, {Hodge}, {Bouwens}, {Mancera Pi{\~n}a}, {Hygate}, {Algera}, {Aravena}, {Bowler}, {da Cunha}, {Dayal}, {Ferrara}, {Herard-Demanche}, {Inami}, {van Leeuwen}, {de Looze}, {Oesch}, {Pallottini}, {Phillips}, {Rybak}, {Schouws}, {Smit}, {Sommovigo}, {Stefanon}, \& {van der Werf}}]{Rowland2024z7}
{Rowland}, L.~E., {Hodge}, J., {Bouwens}, R., {et~al.} 2024, arXiv e-prints, arXiv:2405.06025

\bibitem[{{Sargent} {et~al.}(2014){Sargent}, {Daddi}, {B{\'e}thermin}, {Aussel}, {Magdis}, {Hwang}, {Juneau}, {Elbaz}, \& {da Cunha}}]{Sargent2014}
{Sargent}, M.~T., {Daddi}, E., {B{\'e}thermin}, M., {et~al.} 2014, \apj, 793, 19

\bibitem[{{Schinnerer} {et~al.}(2010){Schinnerer}, {Sargent}, {Bondi}, {Smol{\v c}i{\'c}}, {Datta}, {Carilli}, {Bertoldi}, {Blain}, {Ciliegi}, {Koekemoer}, \& {Scoville}}]{Schinnerer2010}
{Schinnerer}, E., {Sargent}, M.~T., {Bondi}, M., {et~al.} 2010, \apjs, 188, 384

\bibitem[{{Schreiber} {et~al.}(2018){Schreiber}, {Elbaz}, {Pannella}, {Ciesla}, {Wang}, \& {Franco}}]{Schreiber2018Tdust}
{Schreiber}, C., {Elbaz}, D., {Pannella}, M., {et~al.} 2018, \aap, 609, A30

\bibitem[{{Schreiber} {et~al.}(2017){Schreiber}, {Pannella}, {Leiton}, {Elbaz}, {Wang}, {Okumura}, \& {Labb{\'e}}}]{Schreiber2017z4MS}
{Schreiber}, C., {Pannella}, M., {Leiton}, R., {et~al.} 2017, \aap, 599, A134

\bibitem[{{Sillassen} {et~al.}(2024){Sillassen}, {Jin}, {Magdis}, {Daddi}, {Wang}, {Lu}, {Sun}, {Arumugam}, {Liu}, {Brinch}, {D'Eugenio}, {Gobat}, {G{\'o}mez-Guijarro}, {Rich}, {Schinnerer}, {Strazzullo}, {Tan}, {Valentino}, {Wang}, {Xiao}, {Zhou}, {Bl{\'a}nquez-Ses{\'e}}, {Cai}, {Chen}, {Ciesla}, {Dai}, {Delvecchio}, {Elbaz}, {Finoguenov}, {Gao}, {Gu}, {Hale}, {Hao}, {Huang}, {Jarvis}, {Kalita}, {Ke}, {Le Bail}, {Magnelli}, {Shi}, {Vaccari}, {Whittam}, {Yang}, \& {Zhang}}]{Sillassen2024NICE}
{Sillassen}, N.~B., {Jin}, S., {Magdis}, G.~E., {et~al.} 2024, \aap, 690, A55

\bibitem[{{Simpson} {et~al.}(2020){Simpson}, {Smail}, {Dudzevi{\v{c}}i{\={u}}t{\.{e}}}, {Matsuda}, {Hsieh}, {Wang}, {Swinbank}, {Stach}, {An}, {Birkin}, {Ao}, {Bunker}, {Chapman}, {Chen}, {Coppin}, {Ikarashi}, {Ivison}, {Mitsuhashi}, {Saito}, {Umehata}, {Wang}, \& {Zhao}}]{Simpson2020alma}
{Simpson}, J.~M., {Smail}, I., {Dudzevi{\v{c}}i{\={u}}t{\.{e}}}, U., {et~al.} 2020, \mnras, 495, 3409

\bibitem[{{Simpson} {et~al.}(2019){Simpson}, {Smail}, {Swinbank}, {Chapman}, {Chen}, {Geach}, {Matsuda}, {Wang}, {Wang}, {Yang}, {Ao}, {Asquith}, {Bourne}, {Coogan}, {Coppin}, {Gullberg}, {Hine}, {Ho}, {Hwang}, {Ivison}, {Kato}, {Lacaille}, {Lewis}, {Liu}, {Micha{\l}owski}, {Oteo}, {Sawicki}, {Scholtz}, {Smith}, {Thomson}, \& {Wardlow}}]{Simpson2019SCUBA2}
{Simpson}, J.~M., {Smail}, I., {Swinbank}, A.~M., {et~al.} 2019, \apj, 880, 43

\bibitem[{{Simpson} {et~al.}(2014){Simpson}, {Swinbank}, {Smail}, {Alexander}, {Brandt}, {Bertoldi}, {de Breuck}, {Chapman}, {Coppin}, {da Cunha}, {Danielson}, {Dannerbauer}, {Greve}, {Hodge}, {Ivison}, {Karim}, {Knudsen}, {Poggianti}, {Schinnerer}, {Thomson}, {Walter}, {Wardlow}, {Wei{\ss}}, \& {van der Werf}}]{Simpson2014}
{Simpson}, J.~M., {Swinbank}, A.~M., {Smail}, I., {et~al.} 2014, \apj, 788, 125

\bibitem[{{Smail} {et~al.}(2021){Smail}, {Dudzevi{\v{c}}i{\={u}}t{\.{e}}}, {Stach}, {Almaini}, {Birkin}, {Chapman}, {Chen}, {Geach}, {Gullberg}, {Hodge}, {Ikarashi}, {Ivison}, {Scott}, {Simpson}, {Swinbank}, {Thomson}, {Walter}, {Wardlow}, \& {van der Werf}}]{Smail2021Kdrop}
{Smail}, I., {Dudzevi{\v{c}}i{\={u}}t{\.{e}}}, U., {Stach}, S.~M., {et~al.} 2021, \mnras, 502, 3426

\bibitem[{{Smol{\v c}i{\'c}} {et~al.}(2017){Smol{\v c}i{\'c}}, {Novak}, {Bondi}, {Ciliegi}, {Mooley}, {Schinnerer}, {Zamorani}, {Navarrete}, {Bourke}, {Karim}, {Vardoulaki}, {Leslie}, {Delhaize}, {Carilli}, {Myers}, {Baran}, {Delvecchio}, {Miettinen}, {Banfield}, {Balokovi{\'c}}, {Bertoldi}, {Capak}, {Frail}, {Hallinan}, {Hao}, {Herrera Ruiz}, {Horesh}, {Ilbert}, {Intema}, {Jeli{\'c}}, {Kl{\"o}ckner}, {Krpan}, {Kulkarni}, {McCracken}, {Laigle}, {Middleberg}, {Murphy}, {Sargent}, {Scoville}, \& {Sheth}}]{Smolcic2017}
{Smol{\v c}i{\'c}}, V., {Novak}, M., {Bondi}, M., {et~al.} 2017, \aap, 602, A1

\bibitem[{{Strandet} {et~al.}(2017){Strandet}, {Weiss}, {De Breuck}, {Marrone}, {Vieira}, {Aravena}, {Ashby}, {B{\'e}thermin}, {Bothwell}, {Bradford}, {Carlstrom}, {Chapman}, {Cunningham}, {Chen}, {Fassnacht}, {Gonzalez}, {Greve}, {Gullberg}, {Hayward}, {Hezaveh}, {Litke}, {Ma}, {Malkan}, {Menten}, {Miller}, {Murphy}, {Narayanan}, {Phadke}, {Rotermund}, {Spilker}, \& {Sreevani}}]{Strandet2017}
{Strandet}, M.~L., {Weiss}, A., {De Breuck}, C., {et~al.} 2017, \apjl, 842, L15

\bibitem[{{Valentino} {et~al.}(2018){Valentino}, {Magdis}, {Daddi}, {Liu}, {Aravena}, {Bournaud}, {Cibinel}, {Cormier}, {Dickinson}, {Gao}, {Jin}, {Juneau}, {Kartaltepe}, {Lee}, {Madden}, {Puglisi}, {Sanders}, \& {Silverman}}]{Valentino2018CI}
{Valentino}, F., {Magdis}, G.~E., {Daddi}, E., {et~al.} 2018, \apj, 869, 27

\bibitem[{{Vegetti} \& {Koopmans}(2009)}]{Vegetti2009lensing}
{Vegetti}, S. \& {Koopmans}, L.~V.~E. 2009, \mnras, 392, 945

\bibitem[{{Vieira} {et~al.}(2010){Vieira}, {Crawford}, {Switzer}, {Ade}, {Aird}, {Ashby}, {Benson}, {Bleem}, {Brodwin}, {Carlstrom}, {Chang}, {Cho}, {Crites}, {de Haan}, {Dobbs}, {Everett}, {George}, {Gladders}, {Hall}, {Halverson}, {High}, {Holder}, {Holzapfel}, {Hrubes}, {Joy}, {Keisler}, {Knox}, {Lee}, {Leitch}, {Lueker}, {Marrone}, {McIntyre}, {McMahon}, {Mehl}, {Meyer}, {Mohr}, {Montroy}, {Padin}, {Plagge}, {Pryke}, {Reichardt}, {Ruhl}, {Schaffer}, {Shaw}, {Shirokoff}, {Spieler}, {Stalder}, {Staniszewski}, {Stark}, {Vanderlinde}, {Walsh}, {Williamson}, {Yang}, {Zahn}, \& {Zenteno}}]{Vieira2010}
{Vieira}, J.~D., {Crawford}, T.~M., {Switzer}, E.~R., {et~al.} 2010, \apj, 719, 763

\bibitem[{{Vieira} {et~al.}(2013){Vieira}, {Marrone}, {Chapman}, {De Breuck}, {Hezaveh}, {Wei{$\beta$}}, {Aguirre}, {Aird}, {Aravena}, {Ashby}, {Bayliss}, {Benson}, {Biggs}, {Bleem}, {Bock}, {Bothwell}, {Bradford}, {Brodwin}, {Carlstrom}, {Chang}, {Crawford}, {Crites}, {de Haan}, {Dobbs}, {Fomalont}, {Fassnacht}, {George}, {Gladders}, {Gonzalez}, {Greve}, {Gullberg}, {Halverson}, {High}, {Holder}, {Holzapfel}, {Hoover}, {Hrubes}, {Hunter}, {Keisler}, {Lee}, {Leitch}, {Lueker}, {Luong-van}, {Malkan}, {McIntyre}, {McMahon}, {Mehl}, {Menten}, {Meyer}, {Mocanu}, {Murphy}, {Natoli}, {Padin}, {Plagge}, {Reichardt}, {Rest}, {Ruel}, {Ruhl}, {Sharon}, {Schaffer}, {Shaw}, {Shirokoff}, {Spilker}, {Stalder}, {Staniszewski}, {Stark}, {Story}, {Vanderlinde}, {Welikala}, \& {Williamson}}]{Vieira2013}
{Vieira}, J.~D., {Marrone}, D.~P., {Chapman}, S.~C., {et~al.} 2013, \nat, 495, 344

\bibitem[{{Walter} {et~al.}(2012){Walter}, {Decarli}, {Carilli}, {Bertoldi}, {Cox}, {da Cunha}, {Daddi}, {Dickinson}, {Downes}, {Elbaz}, {Ellis}, {Hodge}, {Neri}, {Riechers}, {Weiss}, {Bell}, {Dannerbauer}, {Krips}, {Krumholz}, {Lentati}, {Maiolino}, {Menten}, {Rix}, {Robertson}, {Spinrad}, {Stark}, \& {Stern}}]{Walter2012}
{Walter}, F., {Decarli}, R., {Carilli}, C., {et~al.} 2012, \nat, 486, 233

\bibitem[{{Wang} {et~al.}(2019){Wang}, {Schreiber}, {Elbaz}, {Yoshimura}, {Kohno}, {Shu}, {Yamaguchi}, {Pannella}, {Franco}, {Huang}, {Lim}, \& {Wang}}]{Wang2019Natur}
{Wang}, T., {Schreiber}, C., {Elbaz}, D., {et~al.} 2019, \nat, 572, 211

\bibitem[{{Wardlow} {et~al.}(2011){Wardlow}, {Smail}, {Coppin}, {Alexander}, {Brandt}, {Danielson}, {Luo}, {Swinbank}, {Walter}, {Wei{\ss}}, {Xue}, {Zibetti}, {Bertoldi}, {Biggs}, {Chapman}, {Dannerbauer}, {Dunlop}, {Gawiser}, {Ivison}, {Knudsen}, {Kov{\'a}cs}, {Lacey}, {Menten}, {Padilla}, {Rix}, \& {van der Werf}}]{Wardlow2011}
{Wardlow}, J.~L., {Smail}, I., {Coppin}, K.~E.~K., {et~al.} 2011, \mnras, 415, 1479

\bibitem[{{Witstok} {et~al.}(2022){Witstok}, {Smit}, {Maiolino}, {Kumari}, {Aravena}, {Boogaard}, {Bouwens}, {Carniani}, {Hodge}, {Jones}, {Stefanon}, {van der Werf}, \& {Schouws}}]{Witstok2022-mercurius}
{Witstok}, J., {Smit}, R., {Maiolino}, R., {et~al.} 2022, \mnras, 515, 1751

\bibitem[{{Zavala} {et~al.}(2023){Zavala}, {Buat}, {Casey}, {Finkelstein}, {Burgarella}, {Bagley}, {Ciesla}, {Daddi}, {Dickinson}, {Ferguson}, {Franco}, {Jim{\'e}nez-Andrade}, {Kartaltepe}, {Koekemoer}, {Le Bail}, {Murphy}, {Papovich}, {Tacchella}, {Wilkins}, {Aretxaga}, {Behroozi}, {Champagne}, {Fontana}, {Giavalisco}, {Grazian}, {Grogin}, {Kewley}, {Kocevski}, {Kirkpatrick}, {Lotz}, {Pentericci}, {P{\'e}rez-Gonz{\'a}lez}, {Pirzkal}, {Ravindranath}, {Somerville}, {Trump}, {Yang}, {Yung}, {Almaini}, {Amor{\'\i}n}, {Annunziatella}, {Arrabal Haro}, {Backhaus}, {Barro}, {Bell}, {Bhatawdekar}, {Bisigello}, {Buitrago}, {Calabr{\`o}}, {Castellano}, {Ch{\'a}vez Ortiz}, {Chworowsky}, {Cleri}, {Cohen}, {Cole}, {Cooke}, {Cooper}, {Cooray}, {Costantin}, {Cox}, {Croton}, {Dav{\'e}}, {de La Vega}, {Dekel}, {Elbaz}, {Estrada-Carpenter}, {Fern{\'a}ndez}, {Finkelstein}, {Freundlich}, {Fujimoto}, {Garc{\'\i}a-Argum{\'a}nez}, {Gardner}, {Gawiser}, {G{\'o}mez-Guijarro}, {Guo}, {Hamilton}, {Hathi}, {Holwerda}, {Hirschmann},
  {Huertas-Company}, {Hutchison}, {Iyer}, {Jaskot}, {Jha}, {Jogee}, {Juneau}, {Jung}, {Kassin}, {Kurczynski}, {Larson}, {Leung}, {Long}, {Lucas}, {Magnelli}, {Mantha}, {Matharu}, {McGrath}, {McIntosh}, {Medrano}, {Merlin}, {Mobasher}, {Morales}, {Newman}, {Nicholls}, {Pandya}, {Rafelski}, {Ronayne}, {Rose}, {Ryan}, {Santini}, {Seill{\'e}}, {Shah}, {Shen}, {Simons}, {Snyder}, {Stanway}, {Straughn}, {Teplitz}, {Vanderhoof}, {Vega-Ferrero}, {Wang}, {Weiner}, {Willmer}, {Wuyts}, \& {Ceers Team}}]{Zavala2023photoz}
{Zavala}, J.~A., {Buat}, V., {Casey}, C.~M., {et~al.} 2023, \apjl, 943, L9

\bibitem[{{Zavala} {et~al.}(2018){Zavala}, {Monta{\~n}a}, {Hughes}, {Yun}, {Ivison}, {Valiante}, {Wilner}, {Spilker}, {Aretxaga}, {Eales}, {Avila-Reese}, {Ch{\'a}vez}, {Cooray}, {Dannerbauer}, {Dunlop}, {Dunne}, {G{\'o}mez-Ruiz}, {Micha{\l}owski}, {Narayanan}, {Nayyeri}, {Oteo}, {Rosa Gonz{\'a}lez}, {S{\'a}nchez-Arg{\"u}elles}, {Schloerb}, {Serjeant}, {Smith}, {Terlevich}, {Vega}, {Villalba}, {van der Werf}, {Wilson}, \& {Zeballos}}]{Zavala2017}
{Zavala}, J.~A., {Monta{\~n}a}, A., {Hughes}, D.~H., {et~al.} 2018, Nature Astronomy, 2, 56

\end{thebibliography}

\begin{appendix}

\section{Supporting material}

\begin{table}
    \caption{MIR to radio photometry}
    \centering
    \setlength{\tabcolsep}{2pt}
    \renewcommand\arraystretch{1.5}
    \begin{tabular}{c c c}
    \hline\hline
    Facility &  Band &  Flux/mJy\\
    \hline
    Spitzer/MIPS & 24~$\mu$m &  $0.188\pm0.046$\\
    Herschel/PACS &  100~$\mu$m & $0.01\pm1.57$\\
    Herschel/PACS &  160~$\mu$m & $9.72\pm2.52$\\
    Herschel/SPIRE &  250~$\mu$m & $18.81\pm1.79$ \\
    Herschel/SPIRE &  350~$\mu$m & $28.60\pm2.89$ \\
    Herschel/SPIRE &  500~$\mu$m & $24.66\pm2.06$ \\
    SCUBA-2 &  850~$\mu$m & $9.71\pm0.67$\\
    ALMA$^*$ &  870~$\mu$m & $10.45\pm0.60$\\
    AzTEC &  1.1~mm & $5.55\pm1.29$\\
    MAMBO &  1.2~mm & $4.84\pm0.69$\\
    ALMA &  237.5~GHz & $3.65\pm0.47$\\
    ALMA &  224.8~GHz & $3.41\pm0.41$\\
    ALMA &  145~GHz & $0.62\pm0.07$\\
    ALMA &  100~GHz & $0.071\pm0.016$\\
    VLA & 3~GHz & $(28.6\pm2.8)\times10^{-3}$\\
    VLA & 1.4~GHz & $(91.4\pm10.2)\times10^{-3}$\\
    \hline\hline
    \end{tabular}
    {\\Notes: $^*$ From A3COSMOS catalog \citep{Liu2019A3COSMOS}.}
    \label{tab:photo}
\end{table}

\begin{figure}[tbh]
\centering
\includegraphics[width=0.45\textwidth]{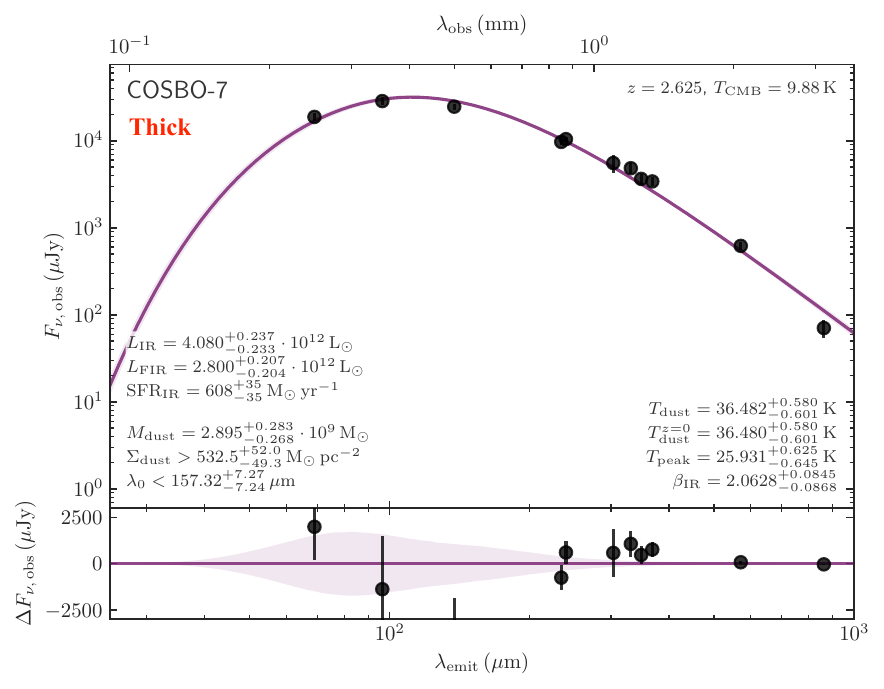}
\includegraphics[width=0.45\textwidth]{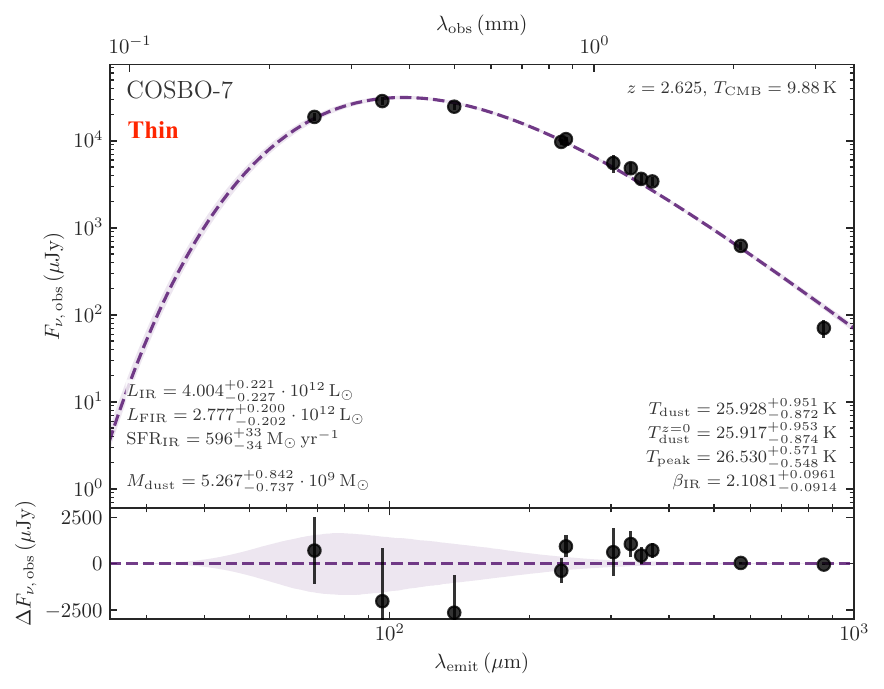}
\caption{
FIR SEDs in optically thick (upper) and thin (bottom) dust models, fitted with \texttt{Mercurius} \citep{Witstok2022-mercurius}.
\label{fig-thick}
}
\end{figure}

\begin{figure}
\centering
\includegraphics[width=0.48\textwidth]{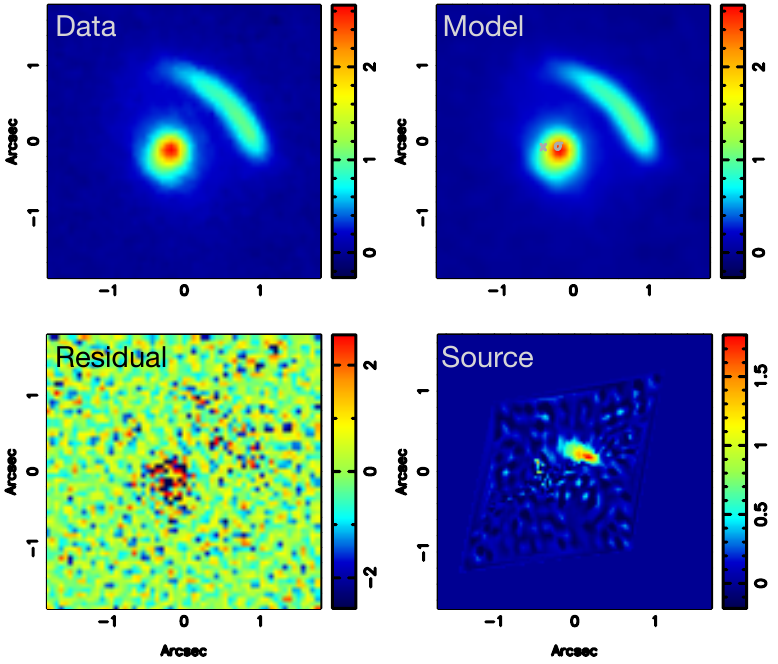}
\caption{
The lens model in the MIRI F770W band. The data is well modeled with a magnification factor $\mu=3.6^{+2.0}_{-0.9}$ using the method {\bf from} \cite{Rizzo2018lensing}. We show the data, model, residual and re-constructed source image, respectively.
\label{fig-lens}
}
\end{figure}

\end{appendix}

\end{document}